\documentclass[aps,prd,amsmath
,eqsecnum            
,preprintnumbers
,nofootinbib         
 ,showpacs          
 ,showkeys          
,twocolumn         
]{revtex4}
\usepackage[dvips]{graphicx}
\usepackage{amsmath}
\usepackage{amsfonts}
\usepackage{amssymb}
\usepackage{longtable}   
\usepackage{color}

\allowdisplaybreaks[4]     

\newcommand{\df}{\ {\overset {\rm def} =}\ }
\newcommand{\dr}[2]{\frac {{\rm d} {#1}} {{\rm d} {#2}}}

\newcommand{\dril}[2]{{{\rm d} {#1}} / {{\rm d} {#2}}}

\newcommand{\llim}[1] {\ {\underset {#1} {\longrightarrow}}\ }

\begin{document}

\title{Cosmological blueshifting may explain the gamma ray bursts}

\author{Andrzej Krasi\'nski}
\affiliation{N. Copernicus Astronomical Centre, Polish Academy of Sciences, \\
Bartycka 18, 00 716 Warszawa, Poland} \email{akr@camk.edu.pl}

\date {}

\begin{abstract}
It is shown that the basic observed properties of the gamma-ray bursts (GRBs)
are accounted for if one assumes that the GRBs arise by blueshifting the
emission radiation of hydrogen and helium generated during the last scattering
epoch. The blueshift generator for a single GRB is a region with a nonconstant bang-time function $t_B(r)$ (described by a Lema\^{\i}tre -- Tolman (L--T) exact solution of Einstein's equations) matched into a homogeneous and isotropic (Friedmann) background. Blueshift visible to the present observer arises \textit{only on those rays that are emitted radially in an L--T region}. The paper presents three L--T models with different Big Bang profiles, adapted for the highest and the lowest end of the GRB frequency range. The models account for: (1) The observed frequency range of the GRBs; (2) Their limited duration; (3) The afterglows; (4) Their hypothetical collimation into narrow jets; (5) The large distances to their sources; (6) The multitude of the observed GRBs. Properties (2), (3) and (6) are accounted for only qualitatively. With a small correction of the parameters of the model, the implied perturbations of the CMB radiation will be consistent with those actually caused by the GRBs. A complete model of the Universe would consist of many L--T regions with different $t_B(r)$ profiles, matched into the same Friedmann background. This paper is meant to be an initial exploration of the possibilities offered by models of this kind; the actual fitting of all parameters to observational results requires fine-tuning of several interconnected variables and is left for a separate study.
\end{abstract}

\maketitle

\section{Motivation and background}\label{intro}

\setcounter{equation}{0}

In the Lema\^{\i}tre \cite{Lema1933} -- Tolman \cite{Tolm1934} (L--T) models, in which the bang-time function $t_B(r)$ is not everywhere constant, \textit{radial} light rays emitted at the Big Bang (BB) can display infinite blueshifts ($z = -1$) to all later observers. This happens when the radial rays are emitted at the generic points of the BB, at which $\dril {t_B} r \neq 0$ \cite{Szek1980,HeLa1984,Kras2014d}. On the other hand, gamma-ray bursts (GRBs) are observed and are believed to originate at large distances from our Galaxy, up to several billion light years \cite{gammainfo}. The question thus arises: could GRBs have been emitted during the last scattering epoch, together with the relic radiation now observed as the cosmic microwave background (CMB), and then blueshifted to gamma-ray frequencies by the mechanism mentioned above?

For the blueshift mechanism to work, the GRBs would have to originate in regions that emerged from a locally delayed BB.\footnote{Regions where the BB occurred
\textit{earlier} than in the background generate shell crossing singularities
\cite{Szek1980,PlKr2006} in addition to blueshifts.} The relic radiation is
emitted a finite time after the BB, at the last-scattering hypersurface (LSH),
so the observed blueshift must be bounded from below, $z \geq z_{\rm LS} > -1$,
where $z_{\rm LS}$ is the blueshift acquired between the LSH and the present
time. The main technical problem to solve is this: can $z_{\rm LS}$ be
sufficiently near to $-1$ that, with the free functions of the L--T model
suitably chosen, the frequencies are blueshifted from the range of the emission
spectra of hydrogen and helium (the only elements present in large amounts
during last scattering) to the gamma-ray range observed today? The present paper answers this question in the positive -- see Sec. \ref{modelfit}.

Section \ref{GRBdata} provides the most basic information on the GRBs. Section
\ref{LTintro} is an introduction to the L--T models, and Sec. \ref{LTnullgeo}
provides information on light propagation in these models. Section
\ref{redshift} discusses the method of calculating and the properties of
redshift in the L--T models. It is shown there that nonradial rays emitted at the BB display infinite redshift to all observers, independent on whether $\dril {t_B} r$ at the BB is zero or not. In Sec. \ref{maxred} the definition of the
extremum-redshift hypersurface (ERH) and the method of determining it are
recalled. In Sec. \ref{Frlim} it is shown how the Friedmann models follow as a
limiting case of the L--T model, and geometric parameters of the now-standard
$\Lambda$CDM model are presented. Section \ref{background} presents the
background Friedmann model used in this paper, it has $\Lambda = 0, k = - 0.4$.

Section \ref{modelfit} presents three L--T models that account for the observed
range of frequencies of the GRBs. Section \ref{shortlive} shows how the third
model qualitatively explains the limited duration of the GRBs. A quantitative
modelling of the duration would require a much higher numerical precision, but
it is shown that the model contains a parameter that can be adapted to the actually observed durations. Section \ref{afterglow} shows how the same Model 3 accounts (qualitatively) for the afterglows of the GRBs.

In Sec. \ref{nonradial}, nonradial rays passing through the L--T region are
discussed. It is shown that the angular diameter of the L--T region as seen by
the present observer in the sky would be $\sim 2$ degrees, and beyond that cone
the CMB radiation would not be perturbed. Thus, with the parameters of the model suitably corrected, the implied perturbations will be hidden within the 1-degree cone of the present resolution of the GRB detectors. The implied pattern of the
perturbations could be observationally tested when the resolution improves.

Section \ref{collimation} explains how the models of Sec. \ref{modelfit} can create the illusion of the collimation of the GRBs into narrow jets, even though they are emitted isotropically by the L--T regions.

Section \ref{distances} shows how the models deal with the large distances to
the sources of the GRBs. If the GRBs are really generated at last scattering,
then the distances to them are even larger than currently believed. However,
with inhomogeneities and blueshifts present, the redshift is not a monotonic
function of distance, and thus fails to be a distance indicator, while the
present estimates of distances to the GRB sources assume a homogeneous
background model and use redshifts of the afterglows for calculating distances. As a by-product it emerges that local blueshifts arise also on nonradial rays, but they are not visible to the present observer: they are overcompensated by redshifts acquired earlier and later along the ray.

Section \ref{multitude} shows that the models used in the paper would allow one to accommodate up to $\sim 10^4$ GRB sources in the sky at the present time.
Decreasing the diameter of the L--T region by a factor $f$ (necessary anyway for other purposes) would increase this number by $f^2$.

Improvements in the model needed to achieve a full quantitative fit are
discussed in Sec. \ref{improve}. Conclusions are summarized in Sec.
\ref{conclu}.

\section{Basic facts about the GRBs}\label{GRBdata}

\setcounter{equation}{0}

The amount of information on the GRBs is enormous \cite{KuZh2015}. In this preliminary study we do not try to interpret all the observational data that are available. Instead, we concentrate on the most characteristic properties of the
GRBs to show how they follow from a simple combined Friedmann/L--T model. The
following properties of the GRBs need to be accounted for \cite{gammainfo}:

(1) Their frequencies extend from $\nu_{\gamma {\rm min}} \approx 0.24 \times
10^{19}$ Hz to $\nu_{\gamma {\rm max}} \approx 1.25 \times 10^{23}$ Hz
\cite{Gold2012}\footnote{Converted from keV to Hz by $\nu = E$/h, where 1 keV = $1.6 \times 10^{-16}$ {\rm J} \cite{unitconver} and h = $6.626 \times 10^{-34}$ J s \cite{constants}. The lowest energy was read out from Figs. 1 and 2 in Ref. \cite{Gold2012} as 10 keV.} (see also Ref. \cite{Grub2014}).

(2) They typically last from less than a second to a few minutes
\cite{gammainfo}, but a few examples are known of GRBs lasting from over two to
about 30 hours \cite{twistGRB}.

(3) For most GRBs, longer-lived and fainter ``afterglows'' at larger wavelengths are observed. It is believed that all GRBs have afterglows, but some of them
were not detected for technical reasons (like the afterglow lasting for too
short a time) \cite{swinafter}.

(4) They are probably focussed into narrow jets.

(5) Nearly all GRBs come from very large distances, from over $10^8$ to several
billion light years.

Currently, there is no universally accepted explanation of origins of the GRBs.
There exist competing attempts at explanation by known astrophysical phenomena
such as gravitational collapse to a black hole, a supernova explosion or a
collision of ultra-dense neutron stars \cite{gammainfo}.

The models presented in Sec. \ref{modelfit} account quantitatively for the
frequency range in property (1), for property (5), qualitatively for properties
(2) and (3), and are consistent with the hypothetical Property (4). References to these properties will be marked by bullets
\textcolor[rgb]{1.00,0.00,0.50}{{\Huge {$\bullet$}}}. Modelling individual GRBs
with full quantitative agreement will require more precise (and much more
time-consuming) numerical fitting of the $t_B(r)$ profiles to the time-profiles
of the GRB frequencies. This is left for future research. The present paper is a proof of existence of L--T models that reproduce the basic GRB properties via suitably adjusted $t_B(r)$ profiles.

\section{The L--T models}\label{LTintro}

\setcounter{equation}{0}

The metric of the L--T models is:
\begin{equation}\label{3.1}
{\rm d} s^2 = {\rm d} t^2 - \frac {{R_{,r}}^2}{1 + 2E(r)}{\rm d} r^2 -
R^2(t,r)({\rm d}\vartheta^2 + \sin^2\vartheta \, {\rm d}\varphi^2),
\end{equation}
where $E(r)$ is an arbitrary function. The source in the Einstein equations is
dust; its (geodesic) velocity field is
\begin{equation}\label{3.2}
u^{\alpha} = {\delta^{\alpha}}_0.
\end{equation}
Because of the property $p = 0$, this model is inadequate for describing times
before the LSH.

The function $R(t, r)$ is determined by
\begin{equation}\label{3.3}
{R_{,t}}^2 = 2E(r) + 2M(r) / R + \frac 1 3\ \Lambda R^2,
\end{equation}
$M(r)$ being another arbitrary function and $\Lambda$ being the cosmological
constant. We consider models with $R,_t > 0$, $E > 0$ and $\Lambda = 0$. The
solution of (\ref{3.3}) is then:
\begin{eqnarray}\label{3.4}
R(t,r) &=& \frac M {2E} (\cosh \eta - 1), \nonumber \\
\sinh \eta - \eta &=& \frac {(2E)^{3/2}} M \left[t - t_B(r)\right],
\end{eqnarray}
where $t_B(r)$ is one more arbitrary function; the BB occurs

\noindent at $t = t_B(r)$. The mass density is
\begin{equation}  \label{3.5}
\kappa \rho = \frac {2{M_{,r}}}{R^2R_{,r}}, \qquad \kappa \df \frac {8\pi G}
{c^2}.
\end{equation}
The $r$-coordinate is chosen so that \cite{Kras2014d}
\begin{equation}\label{3.6}
M = M_0 r^3,
\end{equation}
and $M_0 = 1$ (kept in formulae for dimensional clarity).

The units used in numerical calculations were introduced and justified in Ref.
\cite{Kras2014}. Taking \cite{unitconver}
\begin{equation}\label{3.7}
1\ {\rm pc} = 3.086 \times 10^{13}\ {\rm km}, \quad 1\ {\rm y} = 3.156 \times
10^7\ {\rm s},
\end{equation}
the numerical length unit (NLU) and the numerical time unit (NTU) are defined as follows:
\begin{eqnarray}\label{3.8}
1\ {\rm NTU} &=& 1\ {\rm NLU} = 9.8 \times 10^{10}\ {\rm y} \nonumber \\
&=& 3 \times 10^4\ {\rm Mpc}.
\end{eqnarray}

The L--T models are \textit{generalisations of} (not alternatives to!) the
Friedmann models used in astrophysical cosmology to describe our actual Universe. The relation between these two classes is explained in Sec. \ref{Frlim}.

The following equations that hold in an L--T model with $E \neq 0$ \cite{PlKr2006} will be useful further on:
\begin{eqnarray}
R,_r &=& \left(\frac {M,_r} M - \frac {E,_r} E\right)R \label{3.9} \\
&+& \left[\left(\frac 3 2 \frac {E,_r} E - \frac {M,_r} M\right) \left(t -
t_B\right) - t_{B,r}\right] R,_t, \nonumber \\
R,_{tr} &=& \frac {E,_r} {2E}\ R,_t \label{3.10} \\
&-& \frac M {R^2}\ \left[\left(\frac 3 2 \frac {E,_r} E - \frac {M,_r} M\right)
\left(t - t_B\right) - t_{B,r}\right]. \nonumber
\end{eqnarray}

\section{Light rays in an L--T model}\label{LTnullgeo}

\setcounter{equation}{0}

The general equations defining the tangent vectors $k^{\alpha} = \dril
{x^{\alpha}} {\lambda}$ to geodesics of the metric (\ref{3.1}), with $\lambda$
being the affine parameter, are
\begin{eqnarray}
\dr {k^t} {\lambda} &+& \frac {R,_r R,_{t r}} {1 + 2E}\ \left(k^r\right)^2
\nonumber \\
&+& R R,_t\ \left[\left(k^{\vartheta}\right)^2 + \sin^2\vartheta\
\left(k^{\varphi}\right)^2\right] = 0, \label{4.1} \\
\dr {k^r} {\lambda} &+& 2 \frac {R,_{t r}} {R,_r}\ k^t k^r + \left(\frac {R,_{r
r}} {R,_r} - \frac {E,_r} {1 + 2E}\right)\ \left(k^r\right)^2 \nonumber \\
&-& \frac {(1 + 2E) R} {R,_r}\ \left[\left(k^{\vartheta}\right)^2 +
\sin^2\vartheta\ \left(k^{\varphi}\right)^2\right] = 0, \label{4.2} \\
\dr {k^{\vartheta}} {\lambda} &+& 2 \frac {R,_t} R\ k^t k^{\vartheta} + 2 \frac
{R,_r} R\ k^r k^{\vartheta} - \cos \vartheta\sin \vartheta
\left(k^{\varphi}\right)^2 = 0, \nonumber \\
&& \label{4.3} \\
\dr {k^{\varphi}} {\lambda} &+& 2 \frac {R,_t} R\ k^t k^{\varphi} + 2 \frac
{R,_r} R\ k^r k^{\varphi} + 2 \frac {\cos \vartheta} {\sin \vartheta}\
k^{\vartheta} k^{\varphi} = 0. \nonumber \\
&& \label{4.4}
\end{eqnarray}
The geodesics determined by (\ref{4.1}) -- (\ref{4.4}) are null when
\begin{equation}\label{4.5}
\left(k^t\right)^2 - \frac {{R,_r}^2 \left(k^r\right)^2} {1 + 2E} - R^2
\left[\left(k^{\vartheta}\right)^2 + \sin^2 \vartheta
\left(k^{\varphi}\right)^2\right] = 0.
\end{equation}
Using $R,_t k^t + R,_r k^r = \dril R {\lambda}$ and recalling that
$k^{\vartheta} = \dril {\vartheta} {\lambda}$, the general solution of
(\ref{4.4}) is
\begin{equation}\label{4.6}
R^2 \sin^2 \vartheta k^{\varphi} = J_0,
\end{equation}
where $J_0$ is constant along the geodesic. The special case $J_0 = 0$
corresponds to two situations:

(a) $k^{\varphi} = 0$, i. e. $\varphi$ being constant along the ray, with
$\vartheta$ being, as yet, unspecified, or

(b) $\vartheta = 0$, i.e. the ray proceeding along the axis of symmetry, with
$\varphi$ being undetermined.

Using the above, the general solution of (\ref{4.3}) is
\begin{equation}\label{4.7}
R^4 \left(k^{\vartheta}\right)^2 \sin^2 \vartheta + {J_0}^2 = C^2 \sin^2
\vartheta,
\end{equation}
where $C^2$ is another constant along the geodesic. When $C = 0$, the geodesic is radial. Then $J_0 = 0$ and either (a) $\vartheta = 0$ with $\varphi$ being undetermined or (b) $\vartheta$ has any constant value along the ray and
$\varphi$ is constant in consequence of (\ref{4.6}). When $C = \pm J_0 \neq 0$,
the geodesic remains in the equatorial plane $\vartheta = \pi/2$. Along any {\it single} geodesic, $\vartheta = \pi/2$ can be achieved by a transformation of the $(\vartheta, \varphi)$ coordinates.

For rays with $J_0 \neq 0$, eq. (\ref{4.6}) implies in addition:
\begin{equation}\label{4.8}
k^{\varphi} \equiv \dr {\varphi} {\lambda} \to \infty \qquad {\rm when} \qquad
R \to 0.
\end{equation}
Thus, if such a ray has $|\dril r {\lambda}| < \infty$ at the intersection with
the BB, then $\dril {\varphi} r \llim{t \to t_B} \infty$, i.e. each of these
rays meets the BB being tangent to a surface of constant $r$.

{}From (\ref{4.6}) and (\ref{4.7}) we get
\begin{equation}\label{4.9}
\left(k^{\vartheta}\right)^2 + \sin^2 \vartheta \left(k^{\varphi}\right)^2 = C^2 / R^4,
\end{equation}
and then (\ref{4.5}) becomes
\begin{equation}\label{4.10}
\left(k^t\right)^2 = \frac {{R,_r}^2 \left(k^r\right)^2} {1 + 2E} + \frac {C^2}
{R^2}.
\end{equation}

Using (\ref{4.6}) -- (\ref{4.10}), eqs. (\ref{4.1}) -- (\ref{4.4}) simplify to
\begin{eqnarray}
\dr t {\lambda} &=& k^t, \label{4.11} \\
\dr {k^t} {\lambda} &=& \left[\frac {C^2} {R^2} - \left(k^t\right)^2\right]
\frac {R,_{t r}} {R,_r} - \frac {C^2 R,_t} {R^3}, \label{4.12} \\
k^r &=& \pm \frac {\sqrt{1 + 2E}} {R,_r}\ \sqrt{\left(k^t\right)^2 - \frac {C^2} {R^2}}, \label{4.13} \\
\dr r {\lambda} &=& k^r. \label{4.14}
\end{eqnarray}
The initial data for (\ref{4.11}) -- (\ref{4.14}) are $(t, r) = (t_o, r_o)$ --
the coordinates of the observation point. The values of $\lambda$ will not
appear in the numerical calculations, in all graphs the independent variable will be $r$, so no value for $\lambda(t_o)$ has to be assumed.

The sign in (\ref{4.13}) is $+$ on those segments of the rays, on which $r$ is
increasing, and $-$ where $r$ is decreasing. Care must be taken in numerical
calculations at those points where $r$ changes from increasing to decreasing or
vice versa; there $k^r$ goes through zero and changes sign.

Since the affine parameter is defined up to a constant factor, one more initial
condition is usually assumed; it can be achieved by rescaling $\lambda$:
\begin{equation}
k^t(t_o) = \pm 1 \label{4.15}
\end{equation}
($+$ for future-directed, $-$ for past-directed rays). Why this is convenient
will be seen in Sec. \ref{redshift}.

With (\ref{4.15}) we have from (\ref{4.10}),
\begin{equation}\label{4.16}
C^2 \leq R^2(t_o,r_o) \df {C_o}^2;
\end{equation}
the equality occurs when ${k_o}^r = 0$, i.e. when the ray is tangent to an $r =$ constant sphere at $(t,r) = (t_o, r_o)$.

Given an initial point $(t_o, r_o)$ and an initial direction coded in $C$,
(\ref{4.11}) -- (\ref{4.14}) determine $k^t$ and $k^r$ all along the
ray.\footnote{The initial values of $\vartheta$ and $\varphi$ are irrelevant for (\ref{4.11}) -- (\ref{4.14}) in consequence of (\ref{4.9}): all rays originating on the same sphere $(t, r) = (t_o, r_o)$ with the same value of $C$ have $k^t$
and $k^r$ expressed by the same formulae. The value of $\vartheta_o$ is needed
if we want to know $k^{\vartheta}$ and $k^{\varphi}$ as well. Even then,
$\varphi_o$ is not needed because of spherical symmetry. It becomes needed when
the equations $k^{\alpha} = \dril {x^{\alpha}} {\lambda}$ are integrated to find the path of the ray.} To find $k^{\vartheta}$ and $k^{\varphi}$, one has to
specify $J_0$, then solve (\ref{4.7}) to find $\vartheta(\lambda)$, and finally
calculate $k^{\varphi}(\lambda)$ from (\ref{4.6}). Note that, given $C$, there
is a whole bundle of rays (labelled by $J_0$) that have the same $k^t(\lambda)$
and $k^r(\lambda)$.

In the graphs, only the rays lying in the subspace $\vartheta = \pi/2$ (on which
$J_0 = \pm C$) will be shown. To calculate them, in addition to (\ref{4.11}) --
(\ref{4.14}) one has to integrate (\ref{4.6}), which becomes
\begin{equation}\label{4.17}
\dr {\varphi} {\lambda} = \pm \frac C {R^2},
\end{equation}
with the initial value $\varphi_o = 0$. The ``$-$-rays'' are mirror-images of
those with $+$, and will not appear in the graphs.

On past-directed radial rays, on which $C = 0$, using (\ref{4.10}), the
equations to be integrated numerically are:
\begin{eqnarray}
\dr t {\lambda} &=& k^t, \label{4.18} \\
\dr {k^t} {\lambda} &=& - \left(k^t\right)^2 \frac {R,_{t r}} {R,_r},
\label{4.19} \\
k^r &=& \varepsilon \frac {\sqrt{1 + 2E}} {R,_r}\ k^t, \label{4.20} \\
\dr r {\lambda} &=& k^r, \label{4.21}
\end{eqnarray}
with the initial condition (\ref{4.15}). The sign in (\ref{4.20}) is
$\varepsilon = +1$ on past-inward and future-outward rays and $-1$ on other
rays.

\section{The redshift}\label{redshift}

\setcounter{equation}{0}

The general formula for redshift is \cite{Elli1971}
\begin{equation}\label{5.1}
1  + z = \frac {(u_{\mu} k^{\mu})_e} {(u_{\nu} k^{\nu})_o},
\end{equation}
where $k^{\mu}$ is an affinely parametrised vector field tangent to a ray
connecting the light source and the observer, both comoving with the cosmic
medium. The subscript ``{\it e}'' means ``at the emission event'', ``{\it o}''
means ``at the observation event''.

Consider a ray proceeding from a spacetime point $P_1$ to $P_2$ and then from $P_2$ to $P_3$. Denote the redshifts acquired in the intervals $[P_1, P_2]$, $[P_2, P_3]$ and $[P_1, P_3]$ by $z_{12}$, $z_{23}$ and $z_{13}$, respectively. Then
\begin{equation} \label{5.2}
1 + z_{13} = \left(1 + z_{12}\right) \left(1 + z_{23}\right).
\end{equation}

On past-directed rays, with $u^{\alpha}$ being given by (\ref{3.2}) and using
(\ref{4.15}), we have
\begin{equation}\label{5.3}
1 + z = - k^t.
\end{equation}
On future-directed rays, (\ref{4.15}) and (\ref{5.3}) change to
\begin{equation}\label{5.4}
k^t(t_o) = +1, \qquad 1 + z = k^t.
\end{equation}
The conditions (\ref{5.3}) and (\ref{5.4}) are compatible with (\ref{5.2}).

For nonradial rays, on which $C \neq 0$, the last term in (\ref{4.10}) will go
to infinity when $R \to 0$. Thus, at the BB, independently of whether the second term in (\ref{4.10}) stays finite or becomes infinite
\begin{equation}\label{5.5}
\lim_{R \to 0} \left|k^t\right| \equiv \lim_{R \to 0} z = \infty.
\end{equation}

Now imagine a bundle of rays originating at an off-center event $P_0 \df (t_0,
r_0, \vartheta_0, \varphi_0)$, and going off from $P_0$ to the past in all
directions. Within this bundle, there will be a ray going radially away from the center and another one going radially toward the center; for both of them $C =
0$. Calculating the redshift in a vicinity of the BB for these two rays is a
complicated thing, but power expansions of the redshift formula in $R$
\cite{HeLa1984} and numerical integrations \cite{Kras2014d} both indicate that,
for radial rays, $z \to -1$ as $R \to 0$ when $\dril {t_B} r \neq 0$ at the
intersection of the ray with the BB, and $z \to \infty$ when $\dril {t_B} r =
0$. The value $z = -1$ implies that the observed frequency would be infinite
(which means: very large for rays emitted close to, but later than the BB). The
property $z < 0$ goes by the name \textit{blueshift}, and $z = -1$ means that
the blueshift is infinite. Note that by the very definition of redshift,
\begin{equation}\label{5.6}
1 + z \geq 0
\end{equation}
must always hold.

\section{The extremum-redshift hypersurface}\label{maxred}

\setcounter{equation}{0}

Along a radial ray, $\dril r {\lambda} \equiv k^r \neq 0$ (except possibly at the BB, where $k^t \to 0$, see (\ref{4.20})). Therefore, using (\ref{5.3}) and (\ref{4.20}), Eq. (\ref{4.19}) may be written as \cite{Bond1947}, \cite{PlKr2006}
\begin{equation}\label{6.1}
\frac 1 {1 + z}\ \dr z r = - \varepsilon \frac {R,_{tr}} {\sqrt{1 + 2E}}.
\end{equation}
Thus, $R,_{tr} = 0$ is the locus of extrema of $z$ along radial rays, called the extremum-redshift hypersurface (ERH).

Further in this paper, we will consider an L--T model with $2E = -k r^2$, where
$k =$ constant. The way of determining the ERH in such a model was described in
Ref. \cite{Kras2014d}, and we only copy the results. The value of $\eta$ (defined in (\ref{3.4})) on the ERH is determined by
\begin{equation}\label{6.2}
x^4 + x^3 + k^3 \left(\frac {r t_{B,r}} {4 M_0}\right)^2 = 0,
\end{equation}
where
\begin{equation}\label{6.3}
x \df \sinh^2 (\eta/2).
\end{equation}
Having found (numerically) $x(r)$, and thus also $\eta(r)$ from (\ref{6.3}), we
find $t(r)$ on the ERH from (\ref{3.4}):
\begin{equation}\label{6.4}
t_{\rm ERH}(r) = t_B(r) + \frac {M_0} {(-k)^{3/2}}\ \{\sinh [\eta(r)] -
\eta(r)\}.
\end{equation}

The ERH does not exist along those rays that hit the BB where $t_{B,r} = 0$ and is not determined at $r = 0$ \cite{Kras2014d}, but the limits at $r \to 0$ and at $t_{B,r} \to 0$ of the solution found at $r \times t_{B,r} \neq 0$ may exist. In particular, (\ref{6.2}) -- (\ref{6.4}) imply that $\eta \to 0$ (i.e. $t_{\rm ERH} \to t_B$) at all points where $t_{B,r} \to 0$, and also at $r \to 0$ if $\lim_{r \to 0} t_{B,r}$ is finite. This means that on approaching all those points, the ERH and the BB become arbitrarily close to each other.

Equation (\ref{6.2}) makes no reference to the initial point of the geodesic
arc. Consequently, the ERH is observer-independent. The extremum \textit{value}
of redshift will depend on the initial point, but the \textit{location} of the extremum will not: the extremum of $z$ along a given geodesic will occur always at the same $r$.

Consider a radial ray proceeding to the past from an initial point that lies
later than the ERH. The redshift on it increases from 0 to a local maximum,
achieved at the ERH. Further down the ray, $z$ initially decreases.\footnote{In the models considered further on, the ERH will have the topology of a thick-walled tea cup, and the radial rays will intersect it up to four times. The $z$ along them will thus have up to two local maxima and up to two local minima.} If the ray could continue to the BB, $z$ would either decrease to $-1$ (if $\dril {t_B} r \neq 0$ at the intersection) or increase to infinity (if $\dril {t_B} r = 0$ there). However, the L--T model does not apply at times earlier than the LSH. Can $z$ become, before the ray crosses the LSH, sufficiently negative to shift the optical frequencies to the gamma-ray range? It is shown in Sec. \ref{modelfit} that this is possible when the functions $E(r)$ and $t_B(r)$ are suitably chosen, and the observer is put in the right spacetime region.

\section{The Friedmann limit of the L--T model, the $\Lambda$CDM
model and the last-scattering instant}\label{Frlim}

\setcounter{equation}{0}

The Friedmann limit of (\ref{3.1}) follows when $t_B$ and $E / M^{2/3}$ are
constant. With the coordinate choice (\ref{3.6}) this means $2E = - kr^2$, where $k$ is the Friedmann curvature index. Then (\ref{3.4}) imply $R = r S(t)$, and
\begin{equation}\label{7.1}
{\rm d} s^2 = {\rm d} t^2 - S^2(t) \left[\frac 1 {1 - kr^2} {\rm d} r^2 + r^2
({\rm d}\vartheta^2 + \sin^2\vartheta \, {\rm d}\varphi^2)\right],
\end{equation}
while (\ref{3.4}) reduce to
\begin{eqnarray}\label{7.2}
S(t) &=& \frac {M_0} {(-k)} (\cosh \eta - 1), \nonumber \\
\sinh \eta - \eta &=& \frac {(- k)^{3/2}} {M_0} \left(t - t_B\right).
\end{eqnarray}
Since the Friedmann models are isotropic around every observer world line, every null geodesic is radial, so (\ref{6.1}) can be used. It is then easily
integrated to give
\begin{equation}\label{7.3}
1 + z = S(t_o)/S(t_e),
\end{equation}
where $t_o$ and $t_e$ are the instants of, respectively, the observation and
emission of the light ray (note that (\ref{7.3}) trivially obeys (\ref{5.2})).

The $\Lambda$CDM model, now most often used as the ``standard'' cosmological
model, is a solution of Einstein's equations for the metric (\ref{7.1}) with
dust source and $k = 0 > \Lambda$ \cite{Kras2014}:
\begin{equation}\label{7.4}
S(t) = \left(- \frac {6M_0} {\Lambda}\right)^{1/3} \sinh^{2/3} \left[\frac
{\sqrt {- 3 \Lambda}} 2 \left(t - t_{B\Lambda}\right)\right],
\end{equation}
where $t = t_{B\Lambda}$ is the instant of the BB. It is characterised by the
Hubble parameter,
\begin{equation}\label{7.5}
H_0 = \left.S,_t/S\right|_{t = t_o}
\end{equation}
and two dimensionless constants: the density parameter and the cosmological constant parameter
\begin{equation}\label{7.6}
\left(\Omega_m, \Omega_{\Lambda}\right) \df \frac 1 {3{H_0}^2}
\left.\left(\frac
{8\pi G \rho_0} {c^2}, - \Lambda\right)\right|_{t = t_o}
\end{equation}
that obey $\Omega_m + \Omega_{\Lambda} \equiv 1$; $\rho_0$ is the present mean
mass density in the Universe. The Hubble parameter $H_0$ in (\ref{7.5}) is
related to the Hubble constant ${\cal H}_0$ by
\begin{equation}\label{7.7}
H_0 = {\cal H}_0 / c.
\end{equation}
The current observations imply \cite{Plan2014}
\begin{equation}\label{7.8}
(\Omega_m, \Omega_{\Lambda}) = (0.32, 0.68), \quad {\cal H}_0 = 67.1 {\rm
km}/({\rm s} \times {\rm Mpc}).
\end{equation}
The above, via (\ref{3.7}) -- (\ref{3.8}), (\ref{7.6}) and (\ref{7.7}), leads to
\begin{equation}\label{7.9}
\left[H_0, - \Lambda\right] = \left[6.71\ ({\rm NLU})^{-1}, 91.849164\ ({\rm
NLU})^{-2}\right].
\end{equation}
The age of the $\Lambda$CDM Universe is \cite{Plan2014}
\begin{equation}
T = 13.819 \times 10^9\ {\rm y} = 0.141\ {\rm NTU}. \label{7.10}
\end{equation}

The end of the recombination epoch (the {\it last scattering} instant) occurs when the temperature of the cosmic matter drops below the one needed for ionising the hydrogen atoms. It will be assumed that this temperature is uniquely determined by the local mass density, $\rho_{\rm LS}$. Consequently, it will be the same along every matter world line and in every L--T model. At larger densities, the L--T and $\Lambda$CDM models do not apply.

In the $\Lambda$CDM model the last scattering occurs at the redshift
\cite{Plan2014,Plan2014b}
\begin{equation}\label{7.11}
z_{\rm LS} = 1090.
\end{equation}
Using this value and (\ref{7.4}), one can calculate the corresponding instant
$t_{\rm LS}$ in the $\Lambda$CDM model from \cite{PlKr2006}
\begin{equation}\label{7.12}
1 + z_{\rm LS} = S(T)/S(t_{\rm LS}),
\end{equation}
where $T$ is given by (\ref{7.10}). Namely, using (\ref{7.9}) and (\ref{7.10}) in (\ref{7.4}) one finds
\begin{equation}\label{7.13}
S(T) = 0.51743812113024401.
\end{equation}
Next, using (\ref{7.4}) and (\ref{7.12}) one finds
\begin{equation}\label{7.14}
t_{\rm LS} = \frac 2 {\sqrt{-3 \Lambda}}\ \ln \left(Y + \sqrt{Y^2 + 1}\right),
\end{equation}
where
\begin{equation}\label{7.15}
Y \df (1 + z_{\rm LS})^{-3/2}\ \sinh \left(\tfrac 1 2 \sqrt{- 3 \Lambda}
T\right).
\end{equation}
Then, using (\ref{7.9}), (\ref{7.10}) and (\ref{7.11}), one finds\footnote{See Appendix \ref{ageLS} for a clarification of the possible confusion connected with $z_{\rm LS}$ vs. the age of the Universe at last scattering.}
\begin{eqnarray}\label{7.16}
t_{\rm LS} &=& 4.86905016470083480 \times 10^{-6}\ {\rm NTU} \nonumber \\
&=& 4.7716691614068183 \times 10^5\ {\rm y}.
\end{eqnarray}
The value of $S(t)$ at $t = t_{\rm LS}$ is now calculated from (\ref{7.4}):
\begin{equation}\label{7.17}
S_{\rm LS} = 0.000474278754473001721.
\end{equation}

{}From (\ref{7.9}), (\ref{7.8}) and (\ref{7.6}) using $c = 3 \times 10^8$ m/s
and (\ref{3.7}) -- (\ref{3.8}) we obtain for the present mean mass density\footnote{This value is corrected with respect to Ref. \cite{Kras2014c}.}
\begin{equation}\label{7.18}
\kappa \rho_0 = 3 \Omega_m ({\cal H}_0/c)^2 = 43.223136\ ({\rm NLU})^{-2}.
 \end{equation}
Then, the density at last scattering is
\begin{eqnarray}
\kappa \rho_{\rm LS} &=& \kappa \rho_0 \left[S(T)/S_{\rm LS}\right]^3 \label{7.19} \\
&=& 56.1294161975316 \times 10^9 \ ({\rm NLU})^{-2}. \label{7.20}
\end{eqnarray}

\section{The background model}\label{background}

\setcounter{equation}{0}

Each cosmological model in this paper will consist of a Friedmann background with an L--T island matched into it. The background model will be different from $\Lambda$CDM. For this preliminary study it is preferable to set the cosmological constant zero in order to be able to do exact calculations as much as possible. Our Friedmann background will have the following parameters:
\begin{eqnarray}
\Lambda &=& 0, \label{8.1} \\
k &=& - 0.4, \label{8.2} \\
t_B &\df& t_{\rm Bf} = -0.13945554689046649\ {\rm NTU} \nonumber \\
&\approx& -13.67 \times 10^9\ {\rm years}. \label{8.3}
\end{eqnarray}
The $t_{\rm Bf}$ is taken from Ref. \cite{Kras2014d}, where it was the
asymptotic value of the function $t_B(r)$ in an L--T model that mimicked
accelerating expansion (presented in Ref. \cite{Kras2014}); it differs from
$(-T)$ given in (\ref{7.10}) by $\sim 1.6 \%$. The value of $k$ emerged in
numerical experiments.

For comparing the predictions of our models with the CMB data, we will need the
redshift at last scattering in this background model. The density at last scattering $\rho_{\rm LS}$ given by (\ref{7.20}) must be the same as before, so $S_{\rm LS}$ must be the same, too. This time, however, the present value of $S$ must be calculated for the background model with the parameters (\ref{8.1}) -- (\ref{8.3}). It is, using (\ref{7.2})
\begin{equation}\label{8.4}
\overline{S}_{\rm now} = 0.45180345033414671.
\end{equation}
The resulting redshift between the LSH and $t = 0$ in the background model,
$\overline{S}_{\rm now}/S_{\rm LS}$, comes out to be
\begin{equation}\label{8.5}
1 + z^{\rm b}_{\rm LS} = 952.611615159.
\end{equation}
This differs from (\ref{7.11}) by $\sim 12.7 \%$. The present temperature of the CMB radiation is directly measured, so if (\ref{8.5}) were taken for real, it
would imply that the temperature of the background radiation at emission was
higher by 12.7\% than current knowledge tells us (i.e. $\sim 3380$ K instead of
$\sim 3000$ K). But a more appealing way to cure this discrepancy would be to
change $t_B$ or $k$ (or both) so as to make $\overline{S}_{\rm now}$ larger.
This would require increasing $|t_B|$ or increasing $|k|$ (to produce larger $S$ at a given $t - t_B$ with $k < 0$, larger $|k|$ is needed, see Fig. 17.1 in Ref. \cite{PlKr2006}). However, these changes would have to be accompanied by changes in other parameters of the models presented further on, and such changes would require laborious re-calculations. Therefore, for the beginning, we will show how the present model deals with some properties of the GRBs exactly, and with others only qualitatively, just to explore the existing possibilities. See Sec. \ref{improve} for a discussion of improvements.

\section{Fitting the L--T model to the GRB frequencies (Property (1) in Sec.
\ref{GRBdata})}\label{modelfit}

\setcounter{equation}{0}

As already explained, we assume that the last scattering in any model occurs at
the density given by (\ref{7.20}). The density along a ray is calculated using
(\ref{3.5}), and the value of $z$ at the moment when $\rho = \rho_{\rm LS}$
emerges from (\ref{5.3}) during numerical integration of (\ref{4.18}) --
(\ref{4.21}).

The spectra of the GRBs do not have the black-body forms \cite{GRBrealspectra}, so, if the GRBs arise by the mechanism described below, then different frequencies must be blueshifted independently; see Appendix \ref{hotGRB}.

As the cosmological model we take a Friedmann background, with $k$ and $t_B$ given by (\ref{8.2}) and (\ref{8.3}), into which several L--T regions are matched. Each L--T region has a nonconstant $t_B(r)$, and, being spherically symmetric, defines its own radial directions, independent of the other L--T regions. The BB in such a model consists of flat parts that give rise to the Friedmann background, and spherical humps of delayed BB that evolve into the L--T regions. In this section we consider three examples of single L--T regions surrounded by a Friedmann spacetime. The presence of other L--T regions would not perturb the evolution of any single one, but of course would influence the propagation of light passing through them. The calculations presented here assume that there is no such intervening L--T region between the one that emitted the ray and the observer sitting in the Friedmann background. The perturbations caused by intervening L--T regions may be investigated separately. The central theme of this paper is the existence of the mechanism producing sufficiently strong blueshifts.

We assume that the gamma rays in the GRBs originate as emission radiation of
hydrogen and helium, the only elements present in large amounts during the
recombination epoch, together with the radiation that will later become the CMB. The rays emitted in the Friedmann background and rays emitted non-radially in
the L--T region become redshifted and evolve into the CMB. The rays that are
emitted radially in the L--T region become \textit{blue}shifted and become the
GRBs.

The emission frequencies of hydrogen lie between
\begin{equation}\label{9.1}
\nu_{\rm Hmin} = 4.054 \times 10^{13}\ {\rm Hz},
\end{equation}
corresponding to the wavelength of 7400 nm, and
\begin{equation}\label{9.2}
\nu_{\rm Hmax} = 3.2 \times 10^{15}\ {\rm Hz},
\end{equation}
corresponding to 93.782 nm \cite{Hydrospec}, with the frequency of the most
intense line being
\begin{equation}\label{9.3}
\nu_{\rm Hint} = 2.1876 \times 10^{15}\ {\rm Hz},
\end{equation}
corresponding to $656.2852$ nm. The most intense helium emission lines have
wavelengths between 388 nm and 846 nm, i.e. within the range of the hydrogen
spectrum \cite{helspectr}.

The maximum intensity of the CMB radiation by today is at the frequency of $\sim 200$ GHz \cite{CMBspectrum}. Using (\ref{7.11}) and Wien's law it is easy to calculate that the frequency of maximum intensity at the time of emission must have been $2.18 \times 10^{14}$ Hz. This falls between $\nu_{\rm Hmin}$ and $\nu_{\rm Hmax}$.

The $z$ needed to shift the $\nu_{\rm Hmin}$ to the lowest observed frequency of the gamma-ray bursts \cite{Gold2012},
\begin{equation}\label{9.4}
\nu_{\gamma{\rm min}} \approx 0.24 \times 10^{19}\ {\rm Hz},
\end{equation}
is
\begin{equation}\label{9.5}
1 + z_{\rm max} \approx 1.689 \times 10^{-5}.
\end{equation}
The $z$ needed to shift the $\nu_{\rm Hmax}$ to the maximum recorded cosmic gamma-ray frequency \cite{Gold2012},
\begin{equation}\label{9.6}
\nu_{\gamma {\rm max}} \approx 1.25 \times 10^{23}\ {\rm Hz},
\end{equation}
is
\begin{equation}\label{9.7}
1 + z_{\rm min} \approx 2.56 \times 10^{-8}.
\end{equation}
Consequently, $z$ in the GRB models has to obey
\begin{equation}\label{9.8}
z_{\rm min} < z < z_{\rm max}.
\end{equation}
Actually, if a model that accounts for a given $1 + z$ is already constructed,
then there is no problem with making $1 + z$ larger; the challenge is to account for a sufficiently small $1 + z$. Therefore, we will aim at constructing models
that have $z < z_{\rm min}$ and $z < z_{\rm max}$ at the LSH.

The notation may be confusing here: the $z_{\rm max}$ (so denoted because it is
the greatest value of $z$ that will be needed in GRB models) is associated with
the {\em minimum} values of the emission- and GRB frequencies; similar confusion exists for the $z_{\rm min}$.

The shape of the hump in the BB has to be chosen such that it makes $1 + z$ as
small as indicated by (\ref{9.5}) or (\ref{9.7}), but at the same time the hump
is not too wide (to avoid large perturbations of the CMB) or too high (so that
many humps can be in the field of view of the observer, to account for the
ubiquity of the GRBs). The connection between the shape of the hump and the other quantities will become clear in Secs. \ref{GRBvsCMB} and \ref{multitude}.

The 5-parameter family of profiles shown in Fig. \ref{drawpicture7} emerged in numerical experiments. (For remarks on how this shape was arrived at see Appendix \ref{humpshape}.) Each profile consists of two curved arcs and of a straight line segment joining them. The upper-left arc (shown in thicker line) is a segment of the 4-th degree curve
\begin{equation}\label{9.9}
\frac {r^4} {{B_1}^4} + \frac {\left(t - t_{\rm Bf} - A_0\right)^4} {{B_0}^4} =
1,
\end{equation}
where $t_{\rm Bf}$ is given by (\ref{8.3}). The lower-right arc (also
shown in thicker line) is a segment of the ellipse
\begin{equation}\label{9.10}
\frac {\left(r - B_1 - A_1\right)^2} {{A_1}^2} + \frac {\left(t - t_{\rm Bf} -
A_0\right)^2} {{A_0}^2} = 1.
\end{equation}
The straight line segment passes through the point $(r, t) = (B_1, t_{\rm Bf} + A_0)$ where the full curves would be tangent to each other; the dotted arcs show the parts of the curves that are not included in the hump profile.

The profile in Fig. \ref{drawpicture7} has five parameters: $A_0$, $A_1$, $B_0$, $B_1$, and $x_0$ which determines the slope of the straight segment. The other quantities are determined by these five. Figure \ref{drawpicture7} is not drawn to scale with respect to the values used in actual numerical calculations. In particular, $x_0$ and $A_1$ are greatly exaggerated.\footnote{The straight segment of the BB is invisible in most of the next figures in consequence of $x_0$ being extremely small. It was introduced to keep $\dril {t_B} r$ finite between the curved arcs.}

\begin{figure}[h]
\includegraphics[scale=0.5]{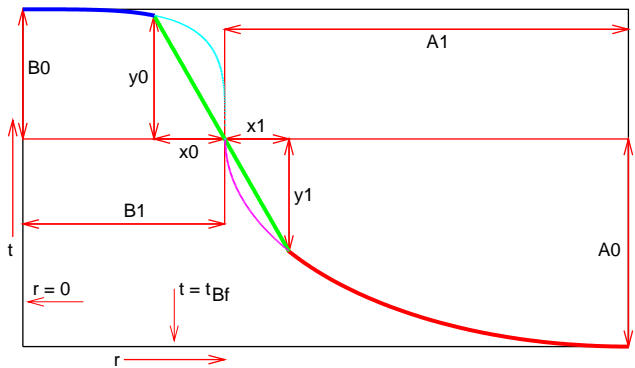}
\caption{Parameters of the bang-time profiles (the profile is drawn not to scale for better readability). See (\ref{9.12}) for the actual values in Model 1 and (\ref{9.19}) for the values in Model 2.}
\label{drawpicture7}
\end{figure}

This exemplary class of profiles is meant to provide a proof of existence of
the blueshifting mechanism. Other profiles may emerge when modelling the actually observed GRBs; see one example in Subsec. \ref{model3}.

The function $E(r)$ obeys
\begin{equation}\label{9.11}
2E/r^2 \df -k = 0.4
\end{equation}
and is the same as in the background model, see Sec. \ref{background}. A general $E$ would have the form $2 E(r) = r^2 [- k + {\cal F}(r)]$, where ${\cal F}(0) = 0$, but otherwise is arbitrary \cite{PlKr2006}.

One more adjustable parameter defines the path of the ray. It is $\Delta t_c$ -- the time-difference between $t_B(0)$ and the instant when the ray intersects the axis $r = 0$. The $\Delta t_c$ determines the coordinate $r$ of the observer who receives the ray at $t = 0$; this way of specifying the initial conditions led to a greater numerical precision.

\subsection{Model 1}

Three models were constructed. Model 1 is adapted to the GRBs of the highest frequency, and satisfies (\ref{9.7}) with some excess. The values of its parameters are
\begin{equation}\label{9.12}
\left(\begin{array}{l}
A_0 \\
B_0 \\
A_1 \\
B_1 \\
x_0\\
\end{array}\right) = \left(\begin{array}{l}
 0.00045\ {\rm NTU} \\
 0.001\ {\rm NTU} \\
 1 \times 10^{-8} \\
 0.03 \\
 2 \times 10^{-13} \\
 \end{array}\right)
\end{equation}
(the last three parameters are dimensionless). The values of $A_0$ and $B_0$
imply the time difference between the maximum of $t_B$ and its flat part 0.00145 NTU = $1.421 \times 10^8$ years ($\sim 0.0103 T$, with $T$ given by
(\ref{7.10})).

The value of $\Delta t_c$ in Model 1 that ensured compatibility with
(\ref{9.7}) is
\begin{equation}\label{9.13}
\Delta t_c = 1.9949248 \times 10^{-5}\ {\rm NTU} \approx 1.955 \times 10^6\ {\rm y}.
\end{equation}
This ray reaches the present time $t = 0$ at
\begin{equation}\label{9.14}
r \df r_{\rm obs1} = 0.76218478306089776.
\end{equation}
With (\ref{9.12}) and (\ref{9.13}), the blueshift is
\begin{equation}\label{9.15}
1 + z_{\rm minb} \approx 2.5191965 \times 10^{-8}
\end{equation}
between the LSH and now. This is slightly better than required by (\ref{9.7}). In this way, the upper limit of GRB frequencies in \textcolor[rgb]{1.00,0.00,0.50}{{\Huge {$\bullet$}} Property (1)} from Sec. \ref{GRBdata} is accounted for. Values of $1 + z$ obeying (\ref{9.7}) could
possibly be obtained with an even smaller height of the BB hump (i.e. with
smaller $A_0 + B_0$). However, no precise optimization was attempted: this model
was meant to be only a proof of existence of the blueshifting mechanism
accounting for the highest GRB frequencies. More precise modelling was done for
Model 3, see Sec. \ref{model3}.

The value (\ref{9.15}) was calculated in two steps, using (\ref{5.2}). First, $1 + z$ between the LSH and the point $P_a$ of coordinates $(r, t) = (0, t_B(0) +
\Delta t_c)$ was calculated; it is
\begin{equation}\label{9.16}
1 + z_{\rm als1} = 9.1144891634 \times 10^{-10}.
\end{equation}
Then, the redshift between $P_a$ and the observer sitting at $t = 0$ was
calculated with the result
\begin{equation}\label{9.17}
1 + z_{\rm oa1} = 27.63947.
\end{equation}
The number in (\ref{9.15}) is the product of the two.

\begin{figure}[h]
\hspace{-6mm}
\includegraphics[scale=0.5]{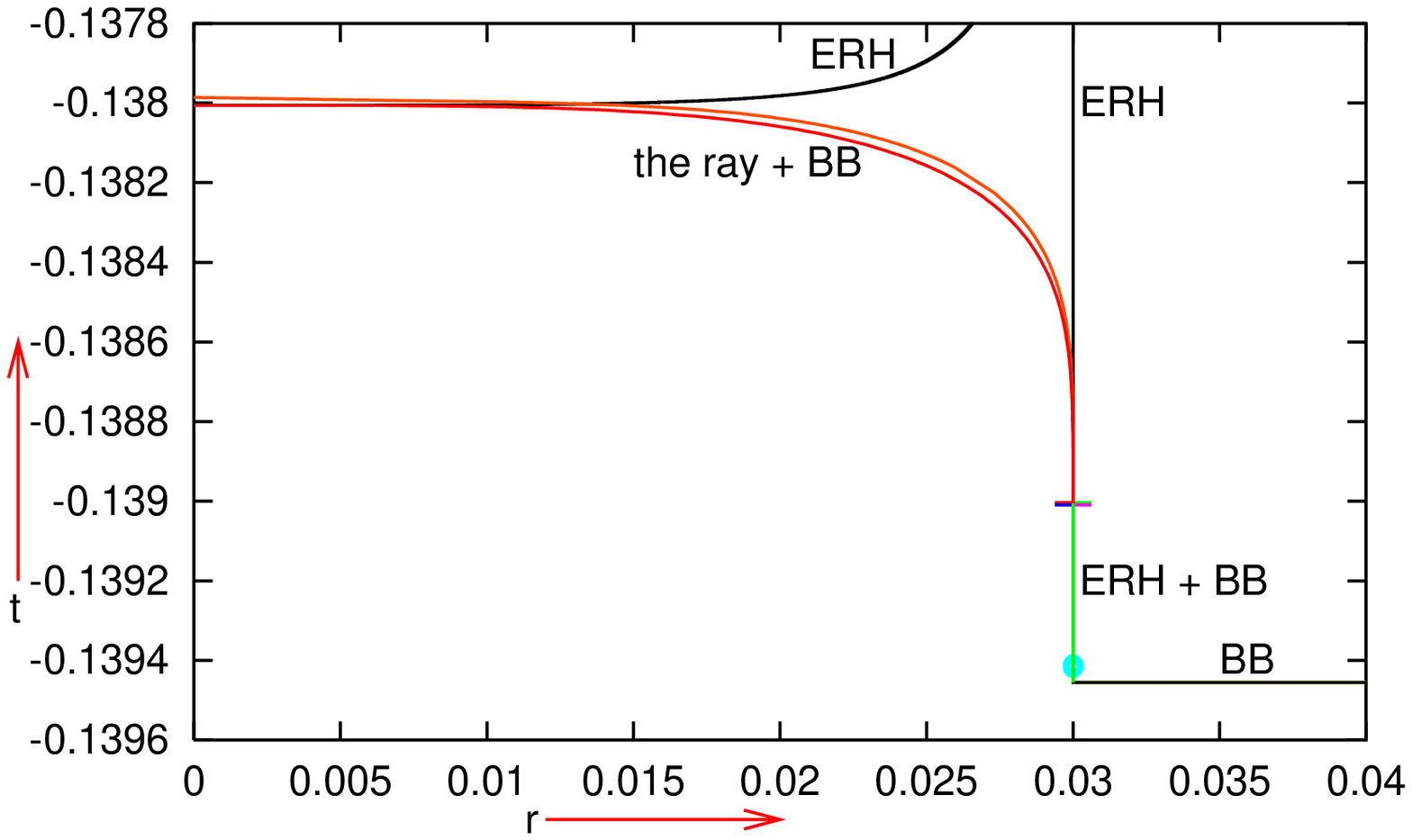}
${ }$ \\[-3.6cm]
\hspace{-1.7cm}
\includegraphics[scale = 0.35]{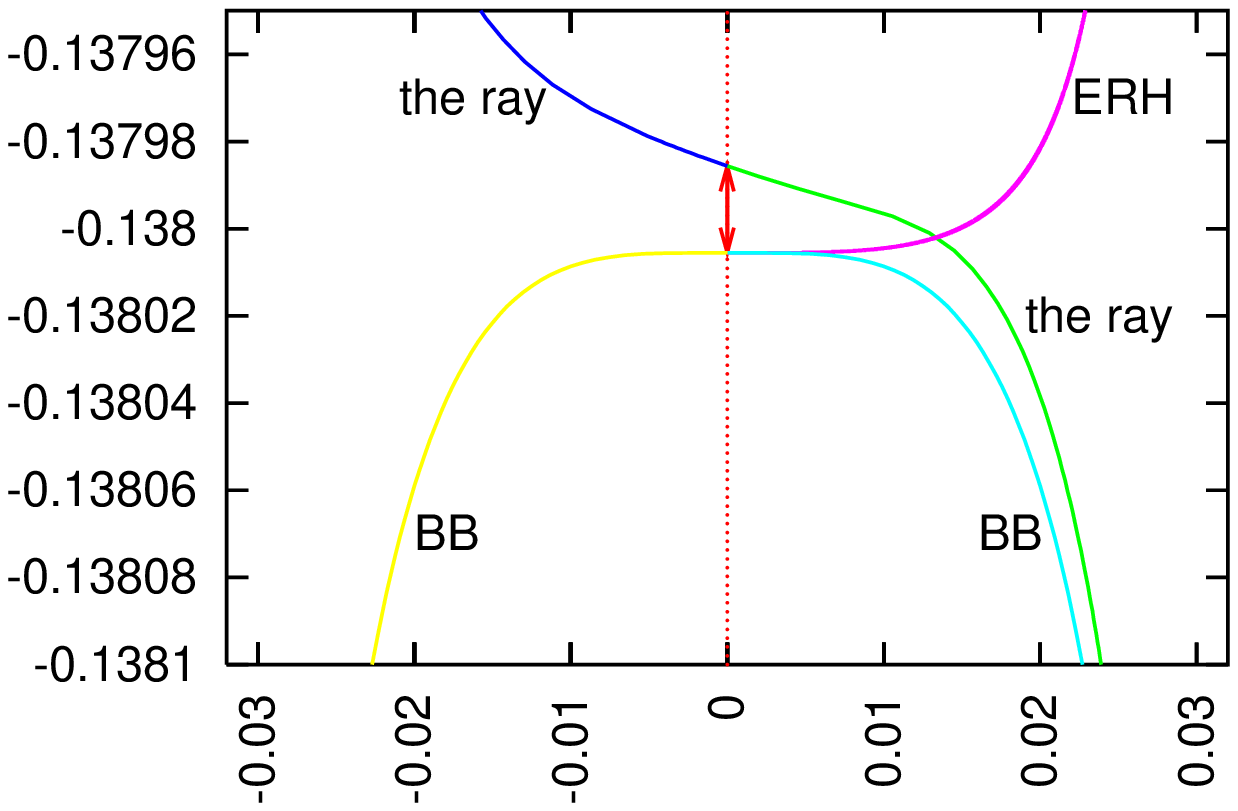}
${ }$ \\[8mm]
\caption{{\bf Main panel:} A radial cross-section through the BB hump in Model
1. It is of the type shown in Fig. \ref{drawpicture7}, with the parameter values
given by (\ref{9.12}). At this scale, the ray defined by (\ref{9.13}) nearly
coincides with the BB. See text for more explanation. {\bf The inset:} A closeup
view of the neighbourhood of the maximum of the BB. The dotted line is $r = 0$.
The difference in $t$ between the ray and the BB at $r = 0$, marked by the
double arrow, is the value of the parameter $\Delta t_c$ given by (\ref{9.13}).}
\label{drawhighfreqray}
\end{figure}

The main panel of Fig. \ref{drawhighfreqray} shows the radial cross-section
through the hump in the BB in Model 1, and the ray defined by (\ref{9.13}). The
inset shows a closeup view of the neighbourhood of $t_B(0)$ (top of the hump).
The LSH and the BB lie so close to each other that they would coincide in all
the figures, so the LSH is not marked.

The line that looks vertical in the main panel is part of the profiles of the
ERH and of the BB that coincide only spuriously, because of the scale of the
figure. It is so close to vertical in consequence of the very small values of
$x_0$ and $A_1$. Another consequence of the small value of $A_1$ is that the
lower (ellipse) arc from Fig. \ref{drawpicture7} is invisible here. The dot near the horizontal part of the BB shows the point where the ray hits the LSH, see
eq. (\ref{d.2}) in Appendix \ref{numtricks} for the exact value of $t$ there
(and compare (\ref{d.3}) to see how close that point is to the BB). The two
(seemingly coincident) horizontal strokes above the dot show the positions of
the ends of the straight segment of the BB. The ERH is spherically symmetric
around the center, so its profile is mirror-symmetric with respect to the $r =
0$ line, but the left part of the profile is omitted in the inset, for clarity
of the picture.

\subsection{Model 2}

Model 2 was obtained from Model 1 by combining a few actions: making the hump lower (by decreasing $A_0$ and $B_0$), changing the slope of its straight part (by increasing or decreasing $x_0$), and replacing (\ref{9.9}) with an analogous curve of still higher degree. Each change in $A_0$ or $x_0$ had to be accompanied by adjusting $A_1$ (to avoid making $x_1$ larger than $A_1$, see Fig. \ref{drawpicture7}) and $\Delta t_c$ (to keep the ray in the range of minimum $1 + z$).

\begin{figure}[h]
\hspace{-6mm}
\includegraphics[scale=0.5]{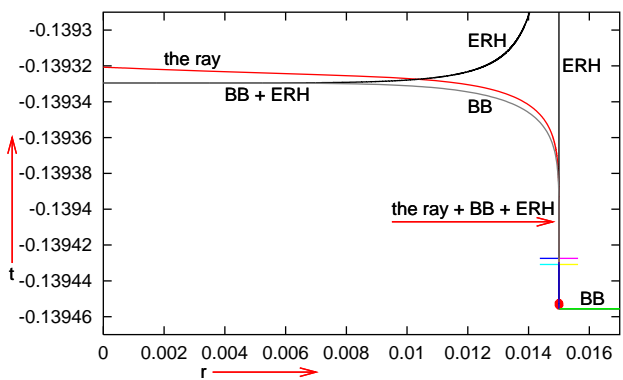}
\caption{This figure is the analogue, for Model 2, of the main panel in Fig.
\ref{drawhighfreqray}. The meaning of the symbols is the same here.}
\label{drawlowfreqray}
\end{figure}

In Model 2, the upper-left arc from Fig. \ref{drawpicture7} is replaced by an
arc of the 6-th degree curve,
\begin{equation}\label{9.18}
\frac {r^6} {{B_1}^6} + \frac {\left(t - t_{\rm Bf} - A_0\right)^6} {{B_0}^6} =
1,
\end{equation}
and its parameters are
\begin{equation} \label{9.19}
\left(\begin{array}{l}
A_0 \\
B_0 \\
A_1 \\
B_1 \\
x_0\\
\end{array}\right) = \left(\begin{array}{l}
 0.000026\ {\rm NTU} \\
 0.0001\ {\rm NTU} \\
 1 \times 10^{-10} \\
 0.015 \\
 2 \times 10^{-13} \\
 \end{array}\right),
\end{equation}
Compared to (\ref{9.12}), $A_0$ is decreased by a factor of $\sim 17$, $B_0$ by
the factor 10, $A_1$ by the factor 100 and $B_1$ by the factor 2. The height of
the hump in the BB is thereby decreased from 0.00145 NTU to 0.000126 NTU, i.e.
by a factor of $\sim 11.5$. The ray that has the smallest $1 + z$ is here
defined by
\begin{equation}\label{9.20}
\Delta t_c = 8.83425 \times 10^{-6}\ {\rm NTU} = 8.657565  \times 10^5\
{\rm y},
\end{equation}
and the blueshift on it is given by
\begin{equation}\label{9.21}
1 + z_{\rm maxb} = 1.36167578 \times 10^{-5},
\end{equation}
being the product of two factors analogous to those in (\ref{9.16}) and
(\ref{9.17}). This time they are
\begin{eqnarray}
1 + z_{\rm als2} &=& 1.07858890707746014 \times 10^{-7}, \label{9.22} \\
1 + z_{\rm oa2} &=& 126.246039921. \label{9.23}
\end{eqnarray}
The $1 + z_{\rm maxb}$ is smaller than the $1 + z_{\rm max}$ given by (\ref{9.5}), and thereby Model 2 accounts for the lower limit of frequencies in
\textcolor[rgb]{1.00,0.00,0.50}{{\Huge {$\bullet$}} Property (1)} from Sec.
\ref{GRBdata}. This ray reaches the present time at the radial coordinate
\begin{equation}\label{9.24}
r \df r_{\rm O2} = 0.88705643159726955.
\end{equation}
The picture corresponding to Fig. \ref{drawhighfreqray} is qualitatively similar here; it is shown in Fig. \ref{drawlowfreqray} to visualise the changes caused
by replacing the curve (\ref{9.9}) with the curve (\ref{9.18}).

\begin{figure}[h]
\hspace{-4mm}
\includegraphics[scale=0.5]{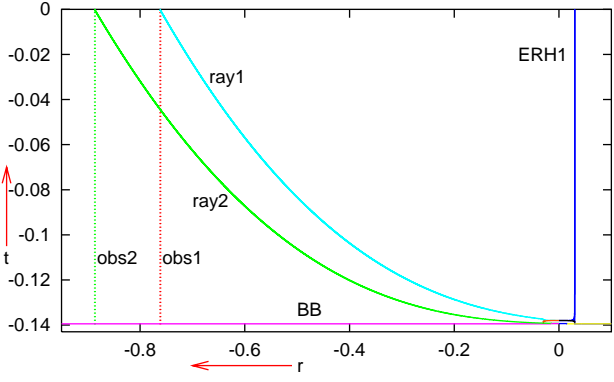}
\caption{Comparison of Models 1 and 2. The right-hand half of the ERH profile is shown only for Model 1.}
\label{drawtruerays}
\end{figure}

\begin{figure}[h]
\includegraphics[scale=0.5]{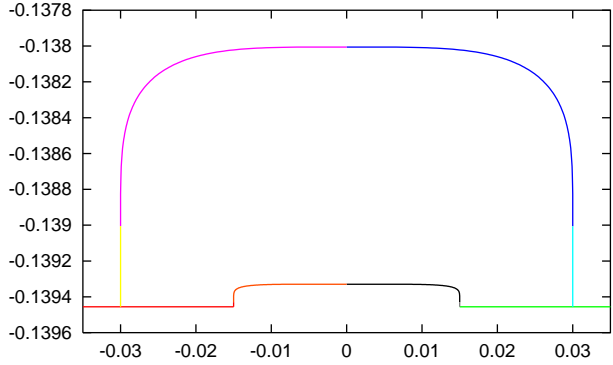}
\caption{Comparison of the BB profile in Model 1 (upper curve, the same as in
Fig. \ref{drawhighfreqray}) with the BB profile in Model 2 (lower curve, the
same as in Fig. \ref{drawlowfreqray}). }
\label{drawcompareprofs}
\end{figure}

Models 1 and 2 are compared in Fig. \ref{drawtruerays}. The BB profiles consist
of the humps from Figs. \ref{drawhighfreqray} and \ref{drawlowfreqray} surrounding the center; further away from the center $t_B$ is constant, and so the geometry is Friedmannian. The present time is $t = 0$, and the flat part of $t_B$ is at $t = t_{\rm Bf}$ given by (\ref{8.3}). The observer lies farther from the hump in Model 2 than in Model 1. The humps are almost invisible here, so they are shown in closeup view in Fig. \ref{drawcompareprofs}.

Some details of the numerical calculations for both models are described in
Appendix \ref{numtricks}.

\subsection{Model 3}\label{model3}

Models 1 and 2 account for the range of frequencies of the GRBs, but, as will be seen in Sec. \ref{shortlive}, do not correctly account for their short duration. Model 3 that will be presented now is a modification of Model 2 made in order to allow us to {\it choose} this time interval.

\begin{figure}
\includegraphics[scale=0.45]{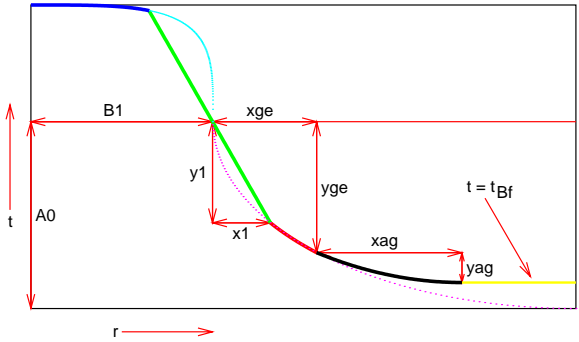}
\caption{The bang-time profile of Model 3 and its parameters. See text for
explanations.} \label{drawpicture8}
\end{figure}

The modified profile of the BB hump is shown in Fig. \ref{drawpicture8} -- again not to scale. Now the ellipse arc is truncated at $r = r_{\rm ge} = B_1 + x_{\rm ge}$ (ge for gamma-burst end), and is replaced by a parabola arc going from $r = r_{\rm ge}$ to $r = r_{\rm ge} + x_{\rm ag}$ (ag for afterglow) achieved at $t = t_{\rm Bf}$ given by (\ref{8.3}). In the special case $x_{\rm ag} = y_{\rm ag} = 0$ the parabola arc is absent, and the ellipse goes over (non-differentiably) into the straight line $t = t_{\rm Bf}$. The parameters $y_{\rm ge}$, $x_{\rm ag}$ and $y_{\rm ag}$ are adjustable, $x_{\rm ge}$ is determined by $y_{\rm ge}$. The net effect compared to Fig. \ref{drawpicture7} is that the whole hump is moved down with respect to the flat part of the BB, by
\begin{equation}\label{9.25}
t_{\rm off} \df A_0 - y_{\rm ge} - y_{\rm ag},
\end{equation}
and the parabola arc interpolates between the ellipse and the straight line. The value of $x_{\rm ge}$ will be given in Sec. \ref{shortlive}, while $x_{\rm ag}
= y_{\rm ag} = 0$ will be assumed in all models.

\section{Accounting for the limited lifetimes of the GRBs (Property (2) in Sec.
\ref{GRBdata})}\label{shortlive}

\setcounter{equation}{0}

\subsection{Adapting the model to the observed durations of
GRBs}\label{adaptdur}

We recall from Sec. \ref{GRBdata} that the observed GRBs typically last from less than a second to a few minutes \cite{gammainfo}. Let us take 10 minutes as the reference value.

If two rays in Model 2 are emitted at $r = 0$ close to the BB, the second one later than the first by $10^{-16}$ NTU ($\approx 5.15$ minutes), then observer O$_2$ sitting at the $r_{\rm O2}$ given by (\ref{9.24}) will receive the second ray $1565.56 \times 10^{-16}$ NTU ($\approx 5.6$ days) later than the first. Consequently, to get a time-interval of 10 minutes $= 1.9414 \times 10^{-16}$ NTU between the received rays at the observer, the time-difference between their emission events at the center would have to be $\sim 1.24 \times 10^{-19}$ NTU. But rays separated by such a short interval of $t$ at $r = 0$, when sent back in time, become indistinguishable for a Fortran90 program at double precision: they (numerically) reach the LSH at the same instant with the same $1 + z$. So, a
much higher numerical accuracy is needed to trace such tiny intervals backward
in time. Lacking this, we will only show that the duration of a GRB in Model 3
can be controlled via the parameter $y_{\rm ge}$.

Numerical experiments with Model 2 showed that the program begins to see the difference between past-directed rays all the way down to the BB when their initial points at $r = 0$ are separated by
\begin{equation}\label{10.1}
\Delta T \approx 1.24 \times 10^{-13}\ {\rm NTU} \quad(\approx 4.4\ {\rm
days}).
\end{equation}
The earlier of the two rays was chosen such that the observer O$_2$ sees it as
(approximately) the first in the gamma-ray frequency range. It crossed the LSH
at
\begin{equation}\label{10.2}
t_{\rm gb2} = -0.13944465218439359\ {\rm NTU}.
\end{equation}
The ray that crossed the center later by $\Delta T$, hit the LSH at
\begin{equation}\label{10.3}
t_{\rm ge2} = -0.13944465708554910\ {\rm NTU} < t_{\rm gb2}.
\end{equation}
(The ray that crosses the LSH earlier reaches the symmetry axis and the observer later, see the figures.)

For the first set of numerical experiments with Model 3, called Case I below, we assumed $x_{\rm ag} = y_{\rm ag} = 0$ and took the difference $t_{\rm ge2} -
t_{\rm Bf}$ as the $t_{\rm off}$ of (\ref{9.25}). Thus
\begin{equation}\label{10.4}
y_{\rm ge3I} = A_0 + t_{\rm Bf} - t_{\rm ge2} = 1.511019508261 \times 10^{-5}\
{\rm NTU}.
\end{equation}

For the ray that defines the beginning of the GR flash in Case I, we took the
one with
\begin{equation}\label{10.5}
\Delta t_c = q_i \df 8.832869 \times 10^{-6}\ {\rm NTU}
\end{equation}
(see Fig. \ref{drawhighfreqray} for the definition of $\Delta t_c$). With
$y_{\rm ge} = y_{\rm ge3I}$ given by (\ref{10.4}), it reached the LSH at
\begin{equation}\label{10.6}
t_{\rm LSH3I} = -0.13945554226395207\ {\rm NTU},
\end{equation}
and the present time $t = 0$ at
\begin{equation}\label{10.7}
r = r_{\rm O3} = 0.89000433423592207.
\end{equation}
Observer O$_3$ residing at $r = r_{\rm O3}$ will be the new reference observer. The redshift on the ray defined by (\ref{10.5}) between the LSH and the O$_3$ world line is
\begin{equation}\label{10.8}
1 + z_{\rm Ib} = 1.67354 \times 10^{-5},
\end{equation}
and satisfies (\ref{9.5}) with a little excess.

For the later rays, $\Delta t_c$ was increased in quanta of
\begin{equation}\label{10.9}
q_t = 10^{-13}\ {\rm NTU}.
\end{equation}
The two rays that had $\Delta t_c = q_i + q_t$ and $\Delta t_c = q_i + 2 q_t$ were still in the gamma-ray frequency range when observed, but the one with $\Delta t_c = q_i + 3 q_t$ was already in the X-ray frequency range, see Table \ref{GRBdura}.

\begin{widetext}
\begin{center}
\begin{table}[h]
\caption{Limits on duration of GRBs in Model 3 with different values of $y_{\rm
ge}$ and $\Delta t_c$ }
\bigskip
\begin{tabular}{|c|c|c|c|c|}
  \hline \hline
\ Case\  & \ Value of $y_{\rm ge}$\ & \ Value of $\Delta t_c$\ & Value of ${\cal
T}$ & $1 + z$  at  O$_3$\\
  \hline \hline
  I & $y_{\rm geI}$ & $\begin{array}{c}
                         q_i + 2 q_t \\
                         q_i + 3 q_t \\
                      \end{array}$ & $\begin{array}{l}
                  9.80519016282838448 \times 10^{-9} \approx 960.90\ {\rm y} \\
                  9.94424450172667973 \times 10^{-9} \approx 974.54\ {\rm y} \\
                                      \end{array}$ & $\begin{array}{c}
                                           1.67815 \times 10^{-5} \\
                                           5.05133912545 \times 10^{-3} \\
                                                   \end{array}$  \\
  \hline
  II & $y_{\rm geII}$ & $\begin{array}{c}
                              q_i + 2 q_t \\
                              q_i + 3 q_t \\
                        \end{array}$ & $\begin{array}{c}
                  9.86541227093332829 \times 10^{-9} \approx 966.81\ {\rm y} \\
                  1.00044488426559296 \times 10^{-8} \approx 980.44\ {\rm y} \\
                                        \end{array}$ & $\begin{array}{c}
                                                 1.676761 \times 10^{-5} \\
                                                 5.051339211 \times 10^{-3} \\
                                                        \end{array}$ \\
  \hline
  III & $y_{\rm geIII}$ & $\begin{array}{c}
                             q_i + 2 q_t \\
                             q_i + 3 q_t \\
                           \end{array}$ & $\begin{array}{c}
                  9.92591136595910132 \times 10^{-9} \approx 972.74\ {\rm y} \\
                  1.00649385349405952 \times 10^{-8} \approx 986.36\ {\rm y} \\
                                           \end{array}$ & $\begin{array}{c}
                                                 1.67476 \times 10^{-5} \\
                                                 5.051339288 \times 10^{-3} \\
                                                           \end{array}$ \\
  \hline \hline
\end{tabular}
\label{GRBdura}
\end{table}
\end{center}
\end{widetext}

In units of NTU, in Case II the value of $y_{\rm ge}$ is
\begin{equation}\label{10.10}
y_{\rm ge3II} = y_{\rm ge3I} + 2 q_t = 1.511019528261 \times 10^{-5},
\end{equation}
and in Case III it is
\begin{equation}\label{10.11}
y_{\rm ge3III} = y_{\rm ge3I} + 4 q_t = 1.511019548261 \times 10^{-5}.
\end{equation}

The values of these and other parameters for all cases are given in Table
\ref{GRBdura}. Assuming that all rays in the table were emitted in the visible
range, in each case the ray with $1 + z \approx 10^{-5}$ is observed in the
gamma-ray range, while the one with $1 + z \approx 10^{-3}$ is observed in the
X-ray range \cite{elecspectr}, i.e. after the end of the GRB. The parameter
${\cal T}$ is the $t$-coordinate of the instant when O$_3$ sees the ray. Values of ${\cal T}$ are given both in NTU (the first one) and in y. The first line of ${\cal T}$ in each case provides the lower limit, the second one provides the upper limit on the duration of the GRB. By this criterion, Table \ref{GRBdura} shows that the duration of the GRB in Model 3 increases when $y_{\rm ge}$ increases, so it is controlled by $y_{\rm ge}$. (In Model 2, which is the limit $y_{\rm ge} = A_0$ of Model 3, the gamma-ray ``burst'' would last more than $10^4$ years.)

The values in Table \ref{GRBdura} are only rough estimates of the GRB duration because they assume that in all three cases the GRB begins at $t = 0$, while in fact the observed frequency might have entered the gamma range earlier. Without a precise identification of the onset of a GRB in a model, no better estimate is possible.

The estimates given here ignore one more problem that is important for detecting the GRBs, namely their intensity. For a GRB to generate a signal in a detector,
it must not only be in the gamma range, but also must be sufficiently strong.
The problem of calculating the intensity of a GRB is left for future
consideration.

\subsection{The time-profile of observed frequencies in case I of Model
3}\label{timeprofile}

We will now follow the sequence of radial rays received by observer O$_3$ in
case I of Model 3, from the (approximately estimated) moment when the BB hump
first appears in her field of view up to the moment (also approximate) when it
disappears. The parameters of several rays emitted at characteristic points of
the LSH are given in Table \ref{tableofrays}. The values of $t$ are given in NTU unless specified otherwise.

\begin{figure}[h]
\includegraphics[scale=0.5]{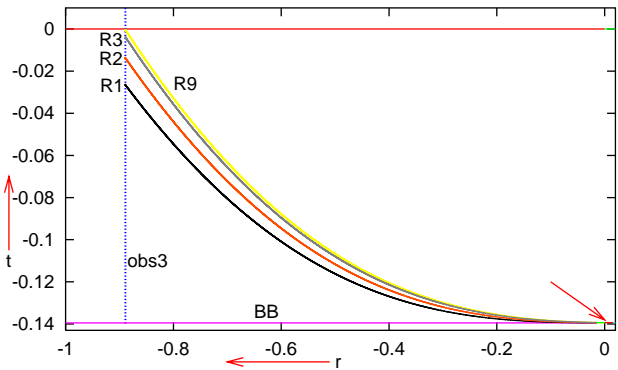}
\caption{Selected rays from Table \ref{tableofrays}. The rays between R3 and R9
are not shown for lack of space. The arrow points to the center of the BB hump. The upper horizontal straight line is the present time. See text for more details. }
\label{trueview}
\end{figure}

\begin{figure}[h]
\includegraphics[scale=0.5]{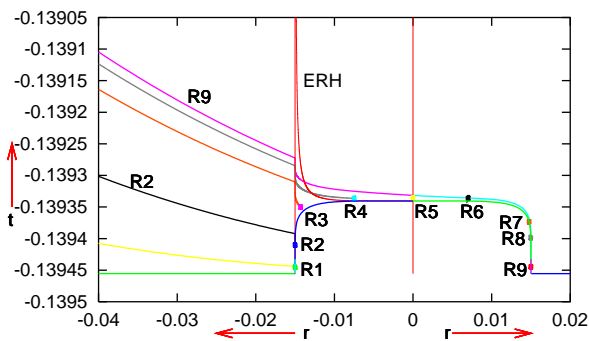}
\caption{Closeup view of the neighbourhood of the BB hump in Fig. \ref{trueview}. Initial points of the rays listed in Table \ref{tableofrays} are marked with dots. The vertical straight line is the axis of symmetry. For rays between R4 and R9 only their initial points are marked, for lack of space.}
\label{trueviewmagni}
\end{figure}

Selected rays are shown in Fig. \ref{trueview} along their whole paths, and in
Fig. \ref{trueviewmagni} in the neigbourhood of the BB. Note how the rays
abruptly change their slopes when crossing the steep wall of the ERH but remain
smooth on intersection with the other branch of the ERH. This is because the
near-vertical part of the BB hump creates a quick rise in the ERH from $t = 0$
to a large value above the upper margin of the figure. This is not a general
property of all ERH, but only of this particular model.

The spurious intersections of different rays in Fig. \ref{trueviewmagni} are
caused by a limited resolution of the figure; in reality the rays do not
intersect.

\begin{widetext}
\begin{center}
\begin{table}[h]
\caption{Parameters of the rays shown in Figs. \ref{trueview} and
\ref{trueviewmagni}}
\bigskip
\begin{tabular}{|c|c|c|c|}
  \hline \hline
  Ray label & $(t, r)$ at the LSH & $t$ at O$_3$ & $1 + z$ at O$_3$ \\
  \hline \hline
  R1 & $\begin{array}{c}
   -0.13944546999999999 \\
   0.0150000000139671984 \\
   \end{array}$ & $\begin{array}{c}
                             -0.02621663\ {\rm NTU} \\
                             \approx 2.569 \times 10^9\ {\rm y\ ago}
                          \end{array}$ & 445.36 \\
  \hline
  R2 & $\begin{array}{c}
   -0.13941000000000001 \\
   0.0149998605746265272 \\
   \end{array}$ & $\begin{array}{c}
                             -0.01342874\ {\rm NTU} \\
                             \approx 1.316 \times 10^9\ {\rm y\ ago}
                          \end{array}$ & 130.439 \\
  \hline
  R3 & $\begin{array}{c}
   -0.13935 \\
   0.0142870667194671501 \\
   \end{array}$ & $\begin{array}{c}
                             -3.4317801 \times 10^{-3}\ {\rm NTU} \\
                             \approx 3.363 \times 10^8\ {\rm y\ ago}
                          \end{array}$ & 75.9982 \\
  \hline
  R4 & $\begin{array}{c}
   -0.13933632431674814 \\
   0.0075\\
   \end{array}$ & $\begin{array}{c}
                             -1.054319 \times 10^{-3}\ {\rm NTU} \\
                             \approx 1.033 \times 10^8\ {\rm y\ ago}
                          \end{array}$ & 92.142 \\
  \hline
  R5 & $\begin{array}{c}
   -0.13933556296232263 \\
   0.0\\
   \end{array}$ & $\begin{array}{c}
                             -5.50955769 \times 10^{-4}\ {\rm NTU} \\
                             \approx 5.399 \times 10^7\ {\rm y\ ago}
                          \end{array}$ & 143.9397 \\
  \hline
  R6 & $\begin{array}{c}
   -0.13933607609165491 \\
   0.00700320078026953671 \\
   \end{array}$ & $\begin{array}{c}
                             -6.484649 \times 10^{-5}\ {\rm NTU} \\
                             \approx 6.3549559 \times 10^6\ {\rm y\ ago}
                          \end{array}$ & 203.5369 \\
  \hline
  R7 & $\begin{array}{c}
   -0.13937350162416995 \\
   0.0147673852333973319 \\
   \end{array}$ & $\begin{array}{c}
                             -6.113304 \times 10^{-5}\ {\rm NTU} \\
                             \approx 5.991 \times 10^6\ {\rm y\ ago}
                          \end{array}$ & 1.6002 \\
  \hline
  R8 & $\begin{array}{c}
   -0.13939833897747514 \\
   0.0149860617889288825 \\
   \end{array}$ & $\begin{array}{c}
                             \approx 2 \times 10^{-8}\ {\rm NTU}\ (*) \\
                             \approx 1960\ {\rm y\ ago} \\
                          \end{array}$ & 0.097862 \\
  \hline
  R9 & $\begin{array}{c}
   -0.13945554226395207 \\
   0.0150000000186084614 \\
   \end{array}$ & 0 (now) & $1.67354 \times 10^{-5}$ \\
  \hline \hline
\end{tabular} \\
 \ \  \\
 (*) -- see remarks in text.
\label{tableofrays}
\end{table}
\end{center}
\end{widetext}

Ray R1 is emitted at the LSH at nearly the same $r$ that is marked with a dot in Fig. \ref{drawlowfreqray}. This is one of the earliest signals from the
neighbourhood of the hump that crosses the world line of observer O$_3$. At a
short initial segment of its path this ray acquires some blueshift. However, on
exit from the ERH the accumulated blueshift is not sufficiently strong to
survive and is completely obliterated by later-acquired redshifts. In the end,
as seen in Table \ref{tableofrays}, ray R1 reaches observer O$_3$ with a
redshift sufficient to shift the lowest-frequency visible light to the
high-frequency segment of microwaves, and the highest-frequency visible light to infrared.

The $1 + z$ at O$_3$ decreases from R1 to R3. This is because, when the emission point of the ray is moved up the hump, the ray travels a longer distance inside the ERH, so it acquires more blueshift. Moreover, the travel time of the ray outside the ERH becomes thereby shorter, so the accumulated redshift becomes smaller.

On rays received after R3, the observer sees increasing $1 + z$ until long past ray R5, emitted at the center of the hump. This happens so because somewhere between R3 and R4 the rays begin to stay within the ERH for shorter and shorter times. At $r = 0$, where R5 has its origin, the ERH is tangent to the BB. Consequently, the initial point of R5 is outside the ERH and the ray builds up positive redshift from the very start.\footnote{The redshift decreases locally on all rays when they pass through the wall of ERH, seen in Fig. \ref{drawtruerays}. But this has little effect on the final value of $1 + z$.} On rays later than R5, $1 + z$ at O$_3$ still increases for a while. This is because with increasing difference in $r$ between the light source and the observer, the redshift acquired outside the ERH outbalances the blueshift acquired inside the ERH.

Somewhere between R6 and R7 $1 + z$ at O$_3$ achieves a maximum and then begins to go down. Between these two rays there is a discontinuity that makes it  extremely difficult (perhaps impossible) to hit, with past-directed rays originating at $r = 0$, the middle of the arc between the points marked R6 and R8. The rays hit the LSH either above that middle point, with $1 + z \approx 1.5$ at O$_3$, or below it, with $1 + z \approx 0.02$ at O$_3$. This is where observer O$_3$ begins to see blueshifts. If this is not a numerical effect, but a reflection of a real discontinuity in the model, then blueshifts appear abruptly, and are initially moderate: $1 + z = 0.02$ takes the visible range into the UV.

A similar instability exists between rays R8 and R9: the change in $1 + z$
measured by O$_3$ from the order of $10^{-2}$ to $10^{-5}$ occurs abruptly. If
both rays are emitted as visible light, then ray R8 is observed still in the UV
range, wile R9 is already in the gamma range. Ray R9 is received by observer
O$_3$ at $t = 0$, i.e., now.

Again, if this is a reflection of a real discontinuity in the model, then the
gamma-ray flash would appear suddenly, without a continuous transition from the
UV range. This seems to agree with the observations of the real GRBs
\cite{gammainfo}.

The asterisk (*) in Table \ref{tableofrays}, in row R8 and column ``$t$ at O$_3$'', indicates that the $t$ at O$_3$ for Ray 8 had to be hand-corrected. The time difference between rays R8 and R9 at O$_3$ is so small that numerical roundoff errors interfere with it. The numerically integrated ray R8 overshot the $r_{\rm O3}$ given by (\ref{10.7}), and the final $t$ reported by the program was greater than zero, which is impossible (if this were true, then R8 and R9 would have to intersect somewhere). So, the $t$ corresponding to $r = r_{\rm O3}$ on ray R8 had to be estimated by comparing the differences in $t$ between R8 and R9 at other values of $r$. Near O$_3$ they were $\sim 2 \times 10^{-8}$ NTU.

For rays received by O$_3$ later than R9, the familiar instability shows up once more:\footnote{The discontinuities in $1 + z$ might be caused by the
non-differentiability of the BB profile at $r = B_1 - x_0$, $r = B_1 + x_1$ and
$r = B_1 + x_{\rm ge}$; see Figs. \ref{drawpicture7} and \ref{drawpicture8}.}
the transition from the gamma-ray range to the X-ray range occurs abruptly when
$\Delta t$ is increased by $10^{-13}$ NTU; see Sec. \ref{adaptdur}. This is the
transition from the GRB to the afterglow.

In addition to the abovementioned discontinuities that are justified by the
properties of the BB profile, there is one more instability that seems to be of
purely numerical character, and which this author was not able to explain or
remove. Namely, when the rays from Fig. \ref{trueviewmagni} are retraced back in time from the observer, they coincide satisfactorily with the original rays only up to the ERH. A numerical instability at the very high and steep wall of ERH
causes that the change of position of the end point (in the past) of the ray is
a discontinuous function of the initial position. The (so far irremovable)
discrepancies between the initial point of the future-directed ray and the end
point of the past-directed ray were of the order of $10^{-6}$ NTU at $r = 0$.

\section{Accounting for the afterglows (Property (3) in Sec.
\ref{GRBdata})}\label{afterglow}

\setcounter{equation}{0}

For this section, we use Case I of Model 3.

\begin{figure}[h]
\begin{center}
\hspace{-5mm}
\includegraphics[scale=0.55]{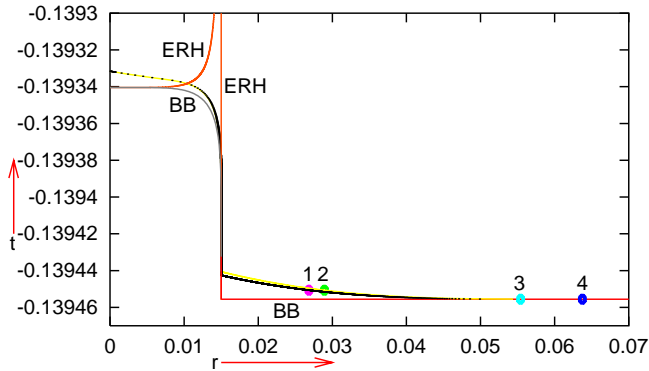}
\caption{Two characteristic rays in the afterglow sector of Model 3. See
explanation in the text.} \label{drawafter}
\end{center}
\end{figure}

In this model, the afterglow appears necessarily. The first ray in the afterglow is the one listed in the second line of Table \ref{GRBdura}. (Figure
\ref{drawafter} does not show it; it would mostly coincide with the BB at this
scale.) If it was emitted in the visible range, then it is observed in the X-ray range.

As time goes by, the observed frequency goes down. For example, the earlier of
the two rays shown in Fig. \ref{drawafter} reaches observer O$_3$ with $z = -0.0008107$, thus its observed frequency is nearly the same as the emitted one. The later ray reaches O$_3$ with $z = 0.598$, which shifts nearly the whole visible range into the infrared. The points marked ``1'' and ``3'' in the figure are at the locations where the earlier ray hits the LSH and the BB, respectively, the points ``2'' and ``4'' show the corresponding events for the later ray. However, this frequency drift occurs at the cosmological time-scale and would not be observable with the current technology \cite{QABC2012}. So, it cannot be responsible for the gradual fading of the afterglow.

The factor that is responsible for the fading of the signal is the intensity of
the received radiation. When it drops below the sensitivity of the detector, the afterglow is, in the technical sense, ended. The question of calculating the intensities of the afterglows in this model is left for a separate investigation.

Model 3 does not account for the duration of an afterglow quantitatively, and
does not contain any parameter that could control this period. As can be seen
from Table \ref{GRBdura}, observer O$_3$ in Case I would begin to see the
afterglow at $t \sim 974.54$ y after now. The later ray in Fig. \ref{drawafter}
is still well within the afterglow, and reaches O$_3$ at $3.5714 \times 10^{-6}$ NTU $\approx 3.5 \times 10^5$ y after now. So, assuming that the intensity of
the ray would still be sufficient for detection, this model-afterglow would
remain visible for observer O$_3$ for nearly 350 000 years, while the
longest-lasting afterglows observed in reality are visible only for several
hundred days \cite{Cenk2011}. Thus \textcolor[rgb]{1.00,0.00,0.50}{{\Huge
{$\bullet$}} Property (3)} from Sec. \ref{GRBdata} is accounted for only
qualitatively.

The blueshifts in the afterglow of Model 3 arise when the rays pass through the
ERH wall seen in Fig. \ref{drawafter}. Consequently, in order to reduce the
duration of the afterglow one should force the ERH to stay nearer to the BB
hump, so that the observed rays begin to bypass the ERH as early as possible.
This should be possible with BB profiles having more parameters, but is not
possible in Model 3. Other BB profiles would also be needed to account for
lower-frequency afterglows. With those profiles, the path of the ray inside the ERH should be suitably short so that the ray would acquire less blueshift.

\section{Nonradial rays}\label{nonradial}

\setcounter{equation}{0}

\subsection{An exemplary nonradial ray hitting the BB}

\begin{figure}[h]
\hspace{-0.4cm}
\includegraphics[scale=0.5]{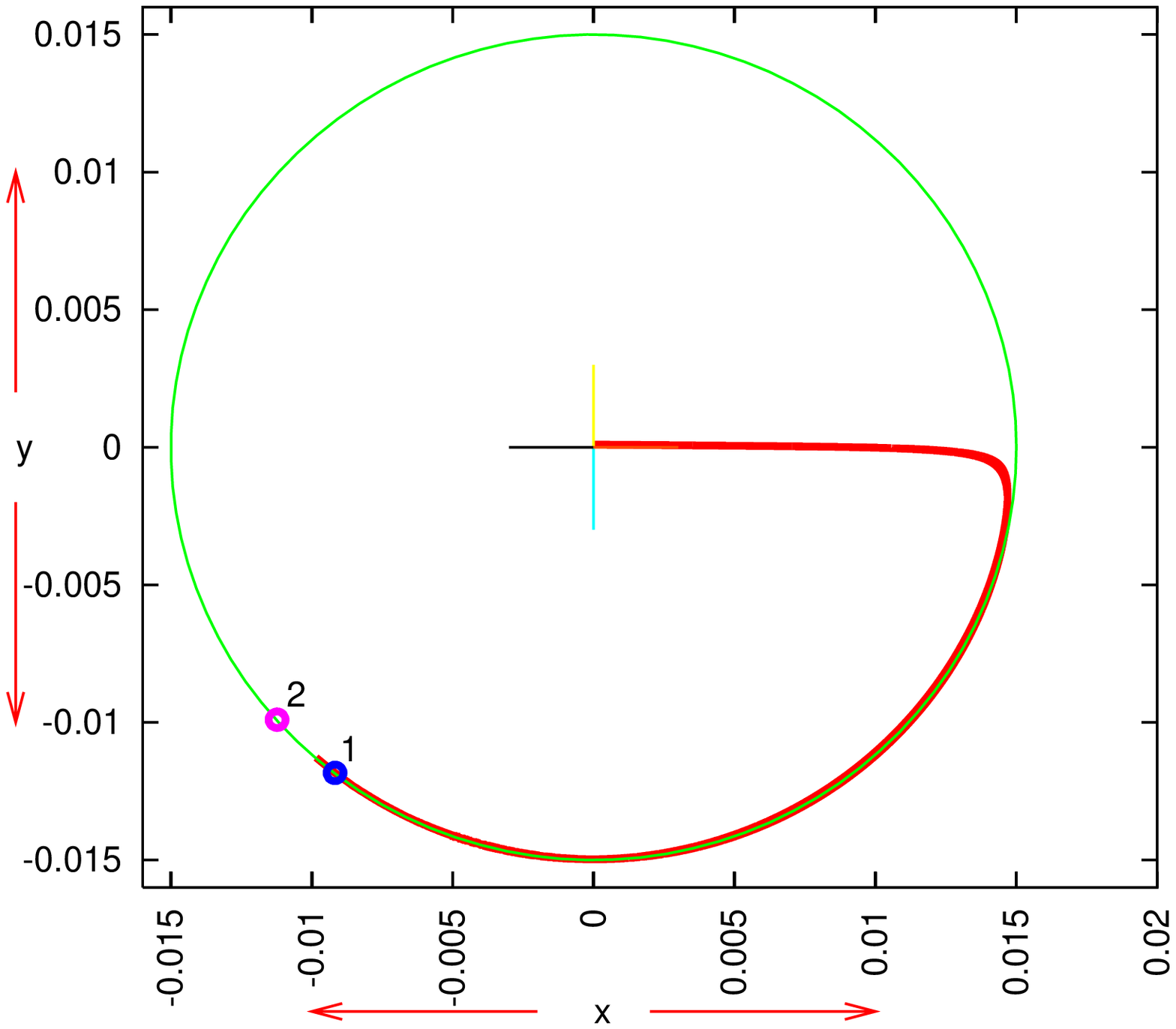}
${ }$ \\[-7cm]
\includegraphics[scale=0.25]{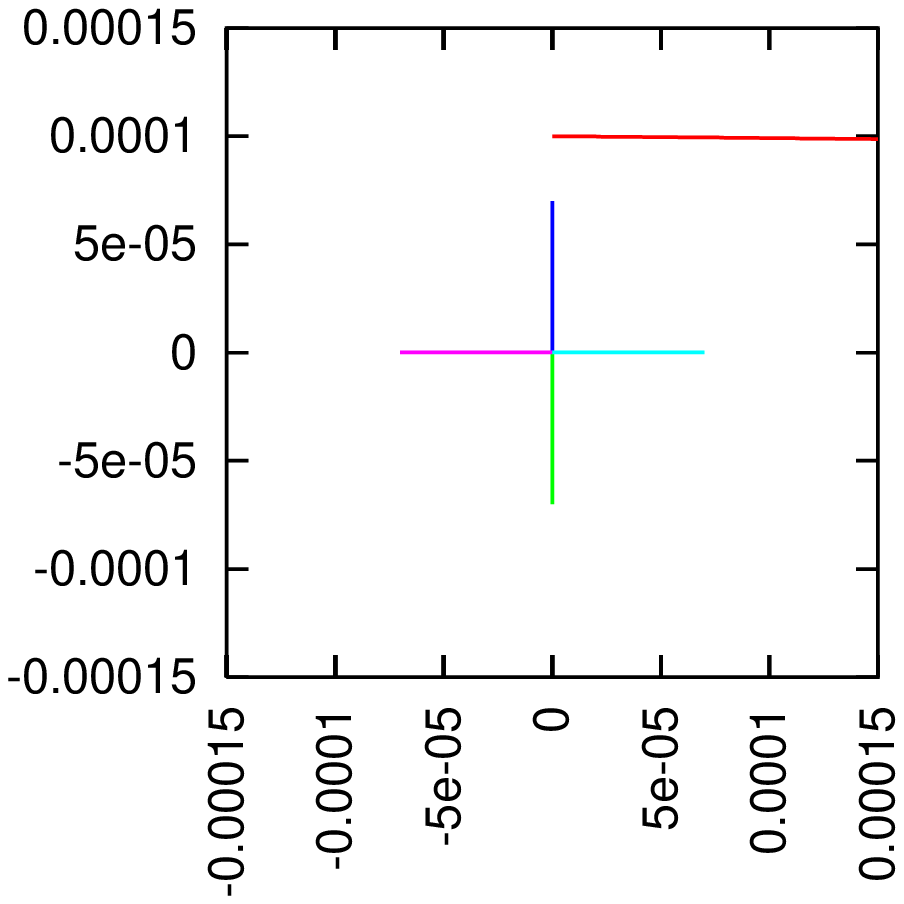}
\vspace{4.5cm} \caption{A nonradial past-directed ray that aims at the BB where
$\dril {t_B} r \neq 0$. It becomes tangent to an $r =$ constant surface at the BB. See text for more explanation.}
\label{illustr}
\end{figure}

As an illustration to the remark under (\ref{4.8}), Fig. \ref{illustr} shows the behaviour of a nonradial ray that hits the BB where $\dril {t_B} r \neq 0$. This is a projection of the ray and of the edge of the BB hump from Fig.
\ref{trueviewmagni} on a surface of constant $t$ along the flow lines of matter
(in comoving coordinates the image is the same at every $t$). For brevity, from
now on we will use just the word ``projection'' to denote this kind of image.
The coordinates in the figure are related to $(r, \varphi)$ by
\begin{equation}\label{12.1}
x = r \cos \varphi, \qquad y = r \sin \varphi.
\end{equation}
The outer edge of the hump is marked with the big circle, its center with the
cross. The ray is emitted toward the past at a point slightly off-center ($r =
0.0001$), in a direction tangent to $r =$ constant, at the same $t =
-0.13933160382638388$ NTU as ray R9. It meets the BB at the steep slope of the
hump. As predicted, it becomes nearly tangent to an $r =$ constant surface
before hitting the BB. The dot \# 1 marks the point where the ray crosses the LSH (with $1 + z = 0.71003236199623432$ relative to the initial point at $r = 0$), the dot \# 2 marks the point where it crosses the BB. Somewhere between these two points the calculation becomes unreliable because of numerical errors (see Appendix \ref{negroot}). The inset is a closeup view on the initial point of the ray; it shows how far off the center the ray begins.

\subsection{Angular size vs. $C$}

\begin{figure}[h]
\includegraphics[scale=0.45]{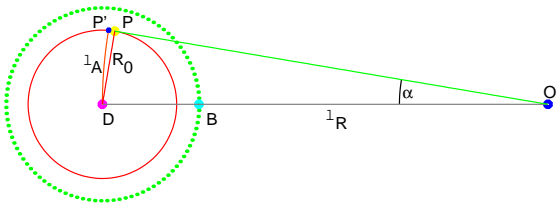}
\caption{Calculating the angular size of an object. See text for explanation.}
\label{anglepic}
\end{figure}

The constant $C$ that first appeared in (\ref{4.7}) is a measure of
non-radialness of a ray. It can be related to the angular size of an object
grazed by the ray. It follows from (\ref{4.13}) using (\ref{5.3}) that at the
point where $\dril r {\lambda} = k^r = 0$ (i.e. where the ray becomes tangent to an $r =$ constant sphere), $C$, $R \df R_0$ and $z \df z_0$ obey
\begin{equation}\label{12.2}
R \df R_0 = |C|/\left(1 + z_0\right).
\end{equation}
The $R_0$ determines the impact parameter, see Fig. \ref{anglepic}.

Figure \ref{anglepic} shows the projection of a ray, similar to Fig.
\ref{illustr}. O is the position of the observer, D is the center of the BB
hump, the dotted circle is the edge of the hump and the smaller solid circle
represents the object whose angular size we wish to calculate. The point P is
the event where (\ref{12.2}) holds, P' is the intersection of the solid circle
with the circle that has the center at O and radius OD, and B is where the
(past-directed) radial ray OD would enter the L--T region. This is an
illustration only; inside the L--T region the projection of a real nonradial ray would not be straight -- see next pictures.

To calculate the angular radius $\alpha$ of the solid circle in Fig.
\ref{anglepic} as seen by the observer at O, two approximations must be made. In a flat space, $\alpha$ would be the length of the arc DP' (call it $\ell_A$)
divided by the length of the straight segment OD (call it $\ell_R$). In the
cosmological context, for $\ell_R$ we can take the angular diameter distance
(ADD). In the L--T model (\ref{3.1}), for an observer at the center of symmetry, ADD is the value of $R$ taken at the location $(t, r)$ of the observed object.
In the Friedmann limit (\ref{7.1}), $R$ becomes $r S(t)$, and every point is a
center. The difficulty in the present case is that the spacetime is not
spherically symmetric around O, and there is no operational definition of the
angular diameter distance in a general cosmological model. So, the two
approximations to make are these:

1. We take $R_0$ for $\ell_A$. In a flat space, the difference between $R_0$ and $\ell_A$ would be $\ell_R (\alpha - \sin \alpha)$, so the relative error at
$\alpha_0 = 1$ degree $= 0.01745$ rad (the actual angular radius of the BB hump
for O, see below) would be $1 - (\sin \alpha_0)/\alpha_0 \approx 5 \times
10^{-5}$.

2. To calculate $\ell_R$ we assume that the ray OD propagates in the Friedmann
background all the way from $r = r_{\rm O3}$ to $r = 0$. The error induced by
this assumption can be estimated in two ways:

(a) As measured by the values of the $r$-coordinate the segment BD is $\sim
0.017$ of OB.

(b) The center $r = 0$ is reached by a ray in Model 3 at
\begin{equation}\label{12.3}
t_{\rm M3} = -0.13932806655865945\ {\rm NTU},
\end{equation}
and in the Friedmann model at
\begin{equation}\label{12.4}
t_{\rm F} = -0.13934135010087942\ {\rm NTU},
\end{equation}
so the difference is
\begin{equation}\label{12.5}
t_{\rm M3} - t_{\rm F} = 0.00001328354221997\ {\rm NTU},
\end{equation}
which is $\sim 0.0095\%$ of $|t_{\rm M3}|$.

Under assumptions 1 and 2, the ADD from O to D is
\begin{equation}\label{12.6}
\ell_R = 3.45912604264922777 \times 10^{-3}\ {\rm NLU};
\end{equation}
see Appendix \ref{der133} for derivation. The angular radii in column 3 of Table \ref{nonradrays} were calculated as $R_0/\ell_R$, where $R_0$ was a by-product in computing the path of the ray.

\subsection{Nonradial rays received by observer O3 in Model 3}

\begin{figure}[h]
\begin{center}
${ }$ \\[-1cm]
\hspace{-4mm}
\includegraphics[scale=0.5]{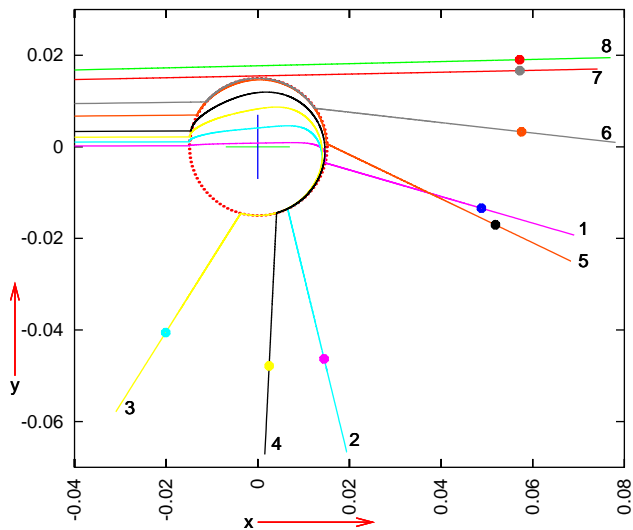}
\caption{View on the rays listed in Table \ref{nonradrays} as they fly over the
BB hump. See explanation in the text. }
 \label{drawpolar}
\end{center}
\end{figure}

\begin{figure}[h]
\begin{center}
\hspace{-7mm}
\includegraphics[scale=0.5]{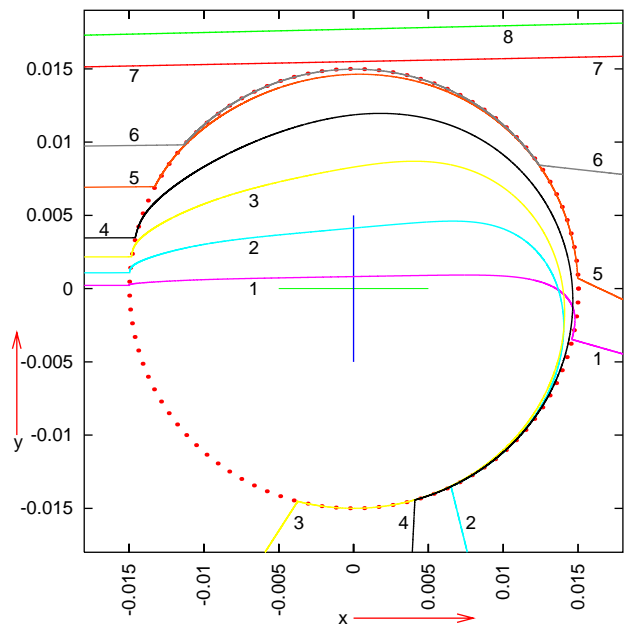}
\caption{A closeup view of the region around $r = 0$ in Fig. \ref{drawpolar}. }
\label{drawpolarmaly}
\end{center}
\end{figure}

\begin{figure}[h]
\begin{center}
\hspace{-7mm}
\includegraphics[scale=0.55]{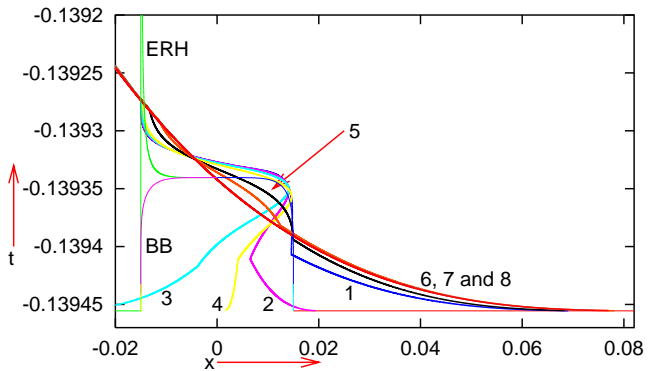}
\caption{Projections of the rays from Fig. \ref{drawpolar} on the radial plane
$y = 0$. See explanation in the text. } \label{drawcross}
\end{center}
\end{figure}

Now we will display several nonradial rays received by observer O3 at the
present time from different directions. We will follow them back in time from the initial point at O3. Table \ref{nonradrays} lists their parameters. The angular radius is calculated as explained above. In Figs. \ref{drawpolar} to \ref{drawz} the horizontal coordinate $x$ is $r \cos \varphi$ and the labels of the rays are the same as in Table \ref{nonradrays}. Note that the redshifts at the LSH increase when the impact parameter increases. The radial ray would have the strong blueshift given by (\ref{10.8}); the changeover to redshift occurs abruptly as soon as the ray ceases to be radial. Ray 6 has larger redshift than the background (\ref{8.5}); this is probably a peculiarity of Model 3 caused by the steep rise of the BB hump profile.

Figure \ref{drawpolar} shows the projection of the rays from Table \ref{nonradrays} in a vicinity of their end points, Fig. \ref{drawpolarmaly} is
a closeup view on the L--T region over the BB hump. The labels of the rays are
printed at their end points; at the left margin of both figures rays 1 -- 8 are
ordered from bottom to top. The dotted circle is the edge of the BB hump, the
cross marks the center $r = 0$. The large dots in Fig. \ref{drawpolar} mark the
points where the rays intersect the LSH. The end points of the rays are where
the numerical calculation determined that the ray crossed the BB.

The rays abruptly change their direction every time they intersect the steep
wall of the ERH (see Figs. \ref{drawhighfreqray} and \ref{drawlowfreqray}). The
change is sharper when the ray is closer to the BB; this happens on the final exit from the ERH. The gentle deflections take place when the rays travel between the two branches of the ERH. (The ERH cannot be shown in these figures: in such a projection the steep wall would coincide with the hump circle, and the other branch projects onto the whole disk inside the circle.)

The angle of deflection of a ray depends on the interval of $t$ that the ray spends between the two branches of the ERH. Ray 1 meets the ERH nearly head-on and does not strongly change direction on first encounter (after the first encounter with the ERH ray 1 stays between the two branches only briefly, similarly to ray R9 in Fig. \ref{trueviewmagni}). However, after the second encounter, it stays between the two branches for a longer interval of $r$ (cp. Fig. \ref{drawlowfreqray}). Then it is close to the BB, where $R$ is small, and is forced to bend around in agreement with (\ref{4.8}).

\begin{center}
\begin{table*}
\caption{Parameters of nonradial rays}
\bigskip
\begin{tabular}{|c|c|c|c|c|c|}
  \hline \hline
  Ray & $C$ & ang. radius ($^{\circ}$) & $r$ at LSH & $1 + z$ at LSH & $r$ at BB
   \\
  \hline \hline
  1 & 0.0001 & 0.0126947 & $5.06068826958865417 \times 10^{-2}$ &
        287.16608259998554 & $7.95216518223493263 \times 10^{-2}$ \\
  \hline
  2 & 0.0005 & 0.064259 & $4.85198622293232865 \times 10^{-2}$ &
      334.76688994090046 & $7.93592114807387117 \times 10^{-2}$\\
  \hline
  3 & 0.001 & 0.1337447 & $4.52659901407877416 \times 10^{-2}$ &
     460.87136026281968 & $7.45576021007735151 \times 10^{-2}$ \\
  \hline
  4 & 0.0016 & 0.226233 & $4.79104086556434161 \times 10^{-2}$ &
     703.10147012965876 & $7.66030728333577382 \times 10^{-2}$ \\
  \hline
  5 & 0.0032 & 0.4743 & $5.45954223781688272 \times 10^{-2}$ &
   945.43487410592388 & $8.45197957881479167 \times 10^{-2}$ \\
  \hline
  6 & 0.0045 & 0.6726575 & $5.76620501069033120 \times 10^{-2}$ &
  970.10190933005345 & $8.60372698978428968 \times 10^{-2}$ \\
  \hline
  7 & 0.007 & 1.00097 & $5.95657733949151447 \times 10^{-2}$ &
  951.91469714961829 & $8.96185695386919057 \times 10^{-2}$ \\
  \hline
  8 & 0.008 & 1.1452 & $6.02258192362461128 \times 10^{-2}$ &
  951.91132098857997 & $8.86836611484453224 \times 10^{-2}$ \\
  \hline \hline
\end{tabular} \\
 \label{nonradrays}
\end{table*}
\end{center}

The other rays meet the ERH at smaller angles than ray 1, so they stay between
the ERH branches for longer times. For rays 2 and 3 this gives the effect that
they are deflected much stronger than ray 1. For ray 4 and the other ones, a
different effect prevails: they stay farther from the BB, so approach the BB
later and stay within its influence for a shorter time, therefore the deflection angle decreases again. Rays 7 and 8, which do not enter the L--T region, do not
feel its influence and propagate just as in a pure Friedmann spacetime.

Figure \ref{drawcross} shows the projections of the rays from Fig.
\ref{drawpolar} on the plane $y = 0$. The intersections of the BB and of the ERH with this plane are also shown (the right part of the ERH profile is suppressed
to avoid clogging the image). In this projection, rays 1 to 4 very nearly
coincide from the left margin of the figure up to the neighbourhood of the right shoulder of the BB hump, and then they split in consequence of the different
deflection angles. Rays 7 and 8 nearly coincide all the way. Ray 6 is clearly
visible only between the ``knot'' where the projections of all the rays
intersect and the right wall of the BB hump, elsewhere it nearly coincides with
7 and 8. Ray 5 is visible on both sides of the knot, and to the right of the BB
hump, where it stays close to 7 and 8.

\subsection{Relation between the GRBs and the CMB}\label{GRBvsCMB}

Redshifts at the LSH listed in Table \ref{nonradrays} must be compared with the
redshift in the background model. The result (\ref{8.5}) was obtained by
substituting numerical values of model parameters into exact formulae. Numerical integration of the null geodesic equation gives, for ray 7, a result differing
by $\sim 0.07\%$:
\begin{equation}\label{12.7}
1 + z_{\rm comp} = 951.91469714961829
\end{equation}
(see Table \ref{nonradrays} for ray 8). Improving the consistency between
(\ref{12.7}) and (\ref{8.5}) would require a higher numerical accuracy, and the
integration time would become prohibitively long. The value (\ref{12.7}) was
found with the step in the affine parameter being $\Delta \lambda = 10^{-9}$ up
to $x = x_b \df -0.03$, and $10^{-13}$ for $x > x_b$.

As Figs. \ref{drawpolarmaly} and \ref{drawcross} show, ray 7 nearly grazes the BB hump. Column 3 in Table \ref{nonradrays} then implies that the area of the sky with redshifts different from the background would fill a cone of angular radius $\sim 1^{\circ}$ for observer O$_3$. This is twice the resolution of current GRB detectors.\footnote{The sources of the GRBs are seen as fuzzy circles of about 1 degree in diameter, private communication from Linda Sparke.} Thus, rays 1 to 5 would be hidden within this circle. Checking for their presence would require aiming a detector at this area while the GRB is still on.

With the redshifts given in column 5 of Table \ref{nonradrays}, all those rays,
if emitted in the visible range, would be seen by the observer in the microwave
range \cite{elecspectr}. It must be recalled here that in Model 3 (and also in
Models 1 and 2) blueshifts visible to the observer are generated only on radial
rays, so the gamma-ray signal, including the afterglow, would have a strictly
point source.

The only reason to worry are rays between 5 and 7, on which the redshift is
different from that of the CMB, and which would be visible around the GRB signal in the present version of Model 3. This shows that Model 3 would need a
modification of its parameters (which is necessary also for other reasons) to
hide the lower-redshift signal within the unresolved patch of gamma radiation.
With that improvement done, the model will not predict perturbations of the
observed CMB larger than the GRBs actually cause. If more-exact measurements in
the future detect some variability within the presently unresolved gamma-ray
dots, then it will be possible to compare it with the model prediction and
improve the model accordingly (or discard it).

\section{Dealing with the collimation of the GRBs (Property (4) in Sec.
\ref{GRBdata})}\label{collimation}

\setcounter{equation}{0}

The blueshifts that account for the GRB frequency range occur on radial rays, and these are emitted isotropically by the L--T region. Thus, there is no real collimation of the GRBs in our models. However, an observer may have an illusion that they are collimated. Namely, as seen from Table \ref{nonradrays}, rays reaching the observer at angle $\beta$ away from radial are seen with large redshifts down to less than $\beta = 0.0127^{\circ}$. So, the GRB source appears to the observer as nearly point-like.

To account for anisotropy and possibly for the collimation, one would have to use an anisotropic model for the BB hump.\footnote{But the collimation of the real GRBs is a hypothesis, not an observationally verified fact \cite{gammainfo}.} A good candidate is the quasi-spherical Szekeres (QSS) metric \cite{Szek1975,Szek1975b}, \cite{PlKr2006}. It contains the L--T model as a subcase, but in general has no symmetry. Nevertheless, it can be matched to any Friedmann background just as well.

Blueshifts visible to the observer are generated in an L--T region only on radial rays, but there is no obvious definition of a radial direction in a QSS spacetime. However, this spacetime contains a flow line of matter that is a natural generalisation of the center of symmetry existing in the L--T models; it was called \textit{origin} in Ref. \cite{HeKr2002}. So, light rays passing through the origin in a QSS spacetime should share some properties with the radial rays of the L--T models. Consequently, it has to be verified what happens with blueshifts when a QSS model is employed (a problem that deserves an investigation independently of the GRB context), and then the question of collimation of the GRBs can be reconsidered.

\section{Accounting for the large distances to the GRB sources (Property (5) in
Sec. \ref{GRBdata})}\label{distances}

\setcounter{equation}{0}

The distances to the GRB sources are inferred by measuring the redshifts in the afterglows \cite{zafterglow} and assuming that the redshift-distance relation that holds in the Robertson -- Walker models applies to them. However, blueshifting renders the redshift-distance relations multivalued (see below and also Ref. \cite{Kras2014c}), so when blueshifts are present along the ray, redshift fails to be an indicator of distance.

In the models hitherto presented, the sources of the GRBs lie close to the Big
Bang, so they are $\sim 13 \times 10^9$ years to the past from the present observers -- still farther than the redshift measurements imply, and in this way \textcolor[rgb]{1.00,0.00,0.50}{{\Huge {$\bullet$}} Property (5)} is accounted for.

\begin{figure}[h]
\begin{center}
\hspace{-7mm}
\includegraphics[scale=0.5]{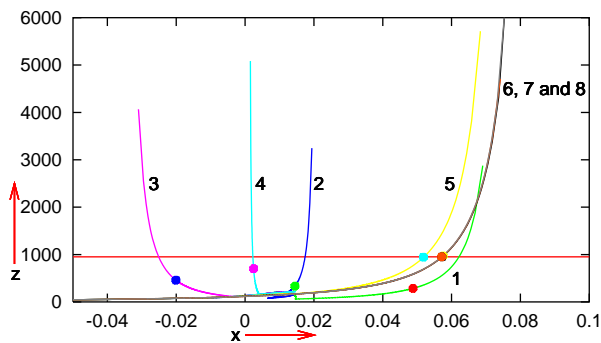}
\caption{The relation between the redshift along the rays from Table \ref{nonradrays} registered by observer O$_3$ and the coordinate $x = r \cos \varphi$ of the light source. See explanation in the text. }
\label{drawz}
\end{center}
\end{figure}

Figure \ref{drawz} shows the relation between the redshift $z$ along the rays from Table \ref{nonradrays} measured by observer O$_3$ (who is placed at $x = - r_{\rm O3}$ given by (\ref{10.7})) and the coordinate $x = r \cos \varphi$ of the light source. The different values of $z$ at which the curves have their end points result from different levels of numerical ``approximation to infinity'' -- on every nonradial ray the redshift between the BB and any observer must be infinite. The background redshift at the LSH given by (\ref{12.7}) is marked with the horizontal line, the large dots mark the value of $z$ between the LSH and O$_3$ on each ray. The redshift profiles on rays 6, 7 and 8 coincide at the scale of the figure, except for the end points. Profile 6 ends at $(r, z) \approx (0.086, 122\ 303.73)$, profiles 7 and 8 end at $(r, z) \approx (0.0896, 299\ 116.45)$ and $(r, z) \approx (0.0887, 166\ 586.96)$, respectively. The graph begins at $x = -0.5$ because at $x < -0.5$ all the curves very nearly coincide and behave in the standard way: $z$ just increases along each ray.

\begin{figure}[h]
\begin{center}
\hspace{-7mm}
\includegraphics[scale=0.5]{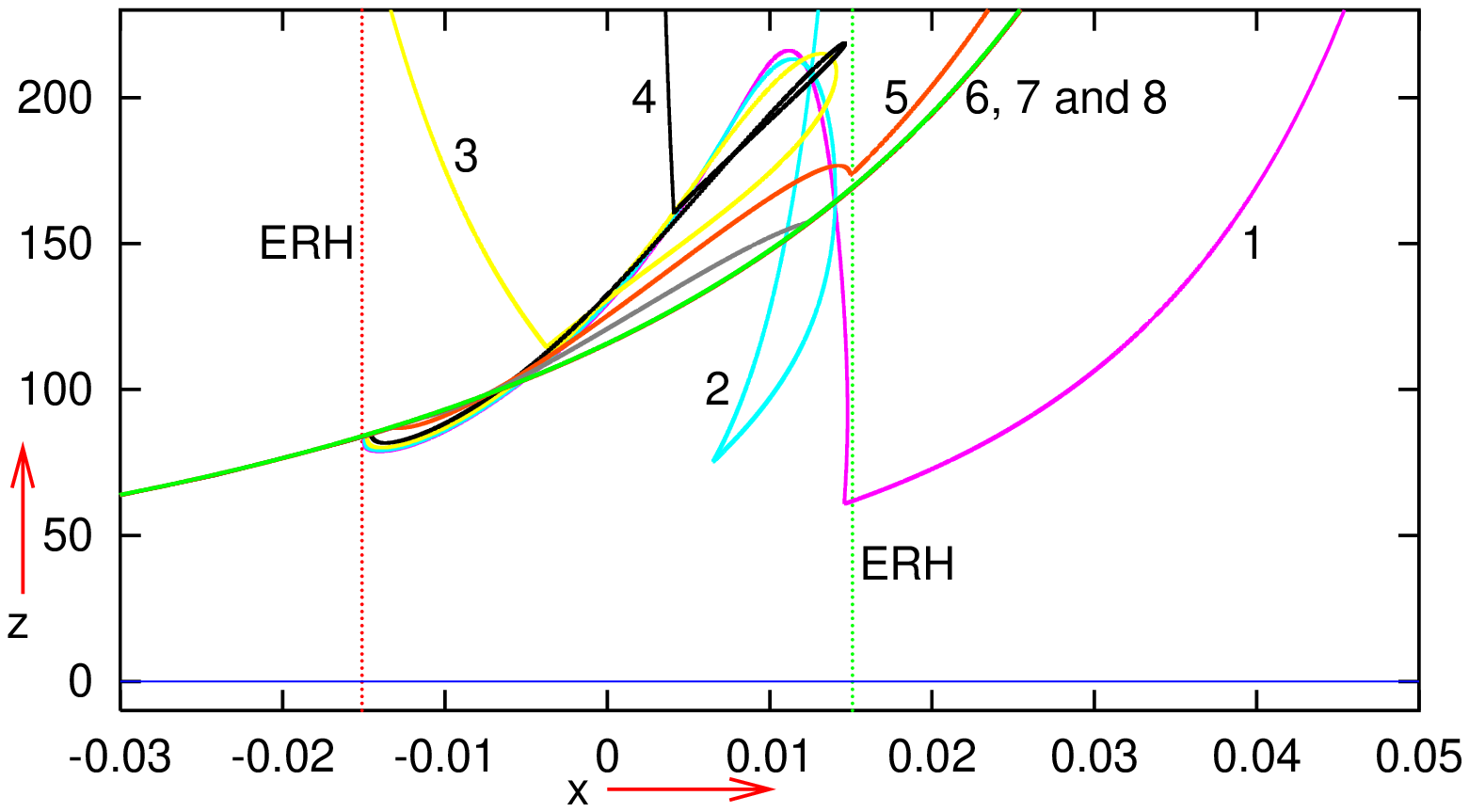}
\includegraphics[scale=0.3]{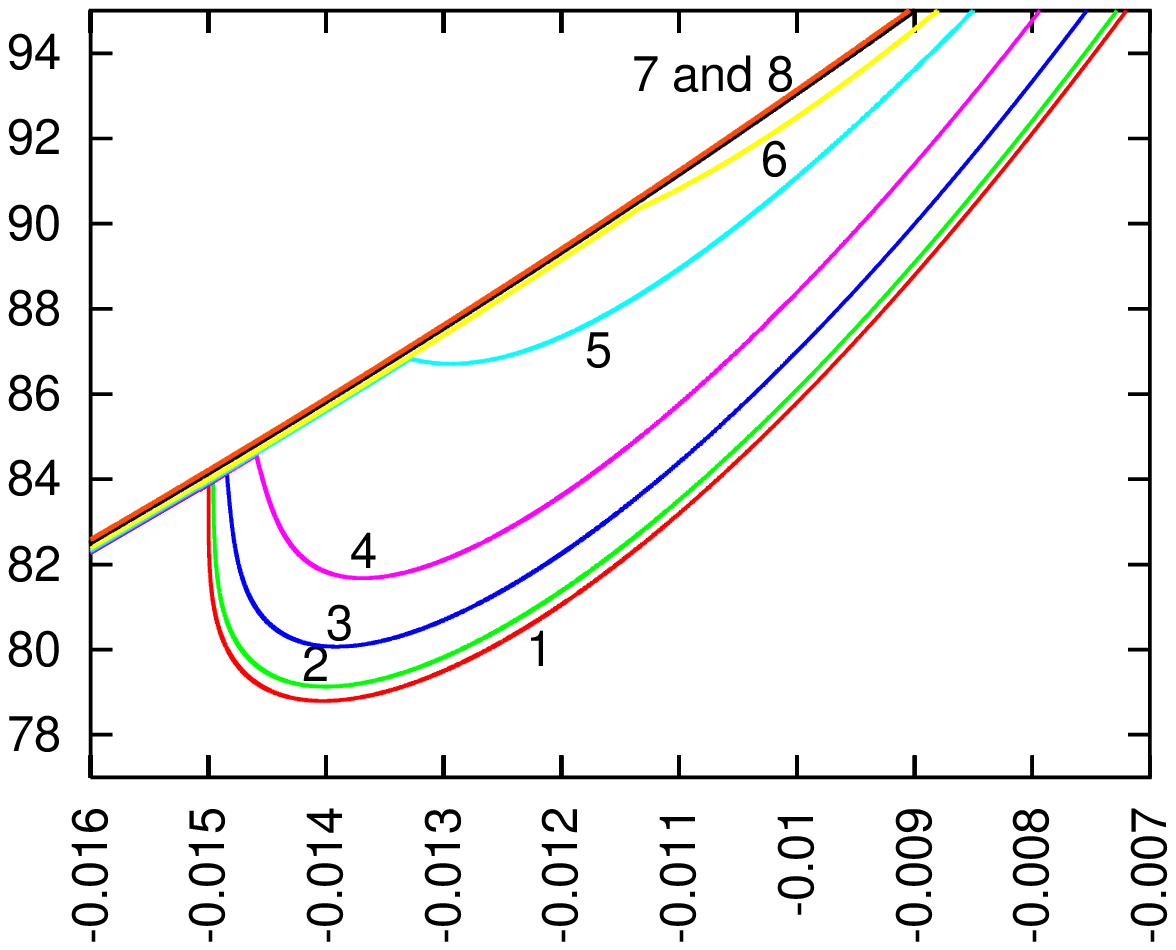}
\includegraphics[scale=0.3]{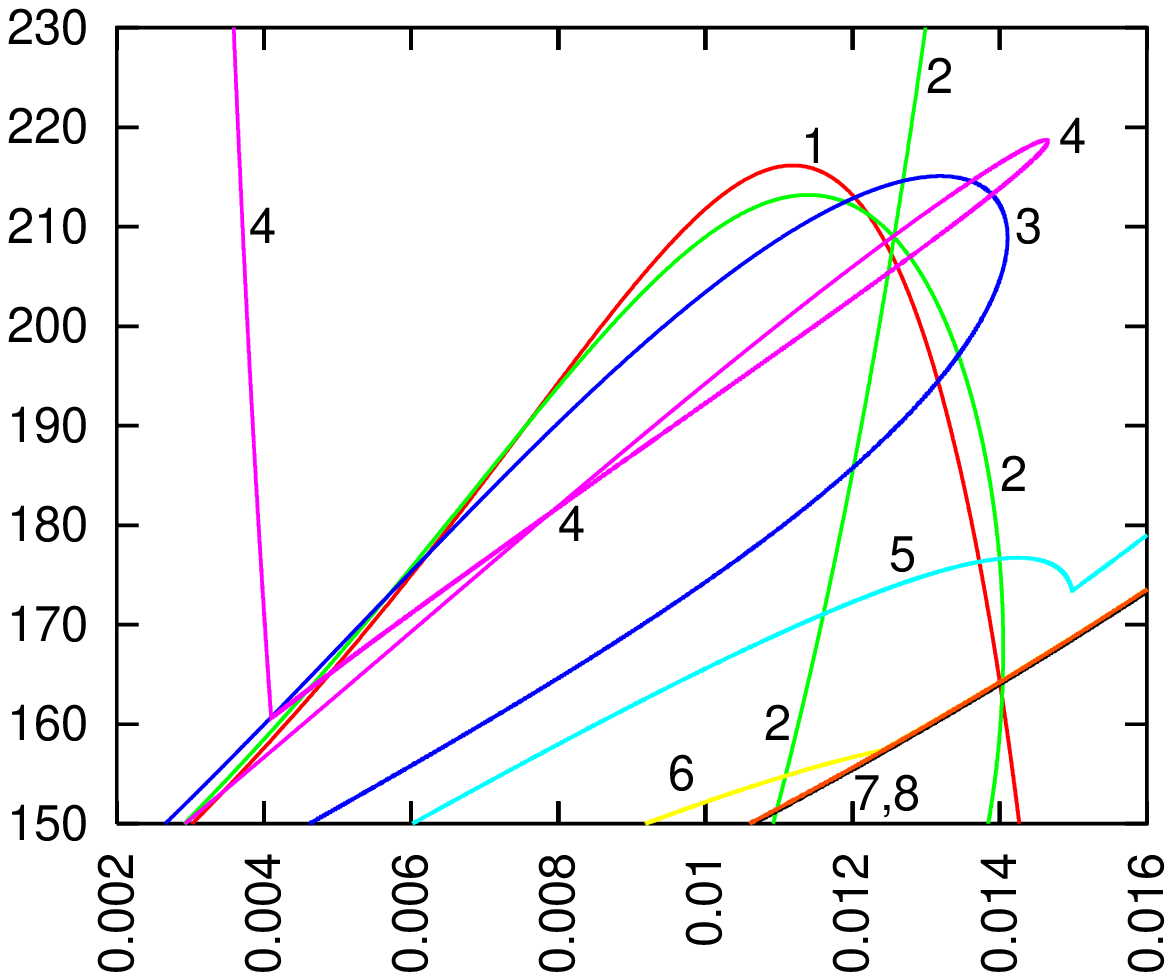}
\caption{Closeup views on the region around $(x, z) = (0, 100)$ in Fig. \ref{drawz}. The redshift is seen to be non-monotonic along the rays inside the ERH. See explanation in the text. }
\label{drawzcent}
\end{center}
\end{figure}

The main panel in Fig. \ref{drawzcent} shows a closeup view on the
neighbourhoood of $(x, z) = (0, 100)$ in Fig. \ref{drawz}, where the redshift
behaves in complicated and untypical ways. The dotted vertical lines mark the
values of $x$ at which the outer ERH intersects the plane $y = 0$.  The two
insets show details of the main panel in two regions where the curves form
particularly complicated knots.

In order to understand these graphs, one has to read them in the correct way. The observer is at $x = - r_{\rm O3}$ given by (\ref{10.7}), beyond the left margin. One has to follow the curves beginning at the position of the observer in the direction of increasing affine parameter. As long as the rays are away from the BB hump (which has its edges at $x \approx \pm 0.015$), all curves very nearly coincide and $z$ increases along them. When a ray goes more than halfway around the hump center (cp. Fig. \ref{drawpolarmaly}), $x$ begins to decrease. This is why each of the curves 2, 3 and 4 turns back at a certain point (and so does curve 1 near its final exit from the ERH, but inconspicuously).

Let us follow ray 2 as an example. Redshift along it increases until the ray
crosses the outer branch of the ERH for the first time. At that crossing, $z$
has a local maximum ($z \approx 84$, see left inset) and begins to decrease. It
decreases until the ray crosses the inner branch of the ERH for the first time
(at $x \approx -0.014$ with $z \approx 79$, left inset). Then $z$ goes up until
the ray crosses the inner ERH for the second time (at $x \approx 0.0118$ with $z \approx 212$, right inset; the graph should be viewed in colour and magnified). From here, $z$ decreases down to the minimum achieved when the ray crosses the outer ERH for the second and last time (at $x \approx 0.007$ with $z \approx 80$, the main panel). Beyond that point, $z$ keeps increasing until the ray crosses the LSH; the coordinates of that point are given in Table \ref{nonradrays}; see also Fig. \ref{drawz}.

Two local maxima and two local minima can be seen along each of rays 1 -- 5. Ray 6 does not intersect the ERH, so has no local extrema, but redshift along it shows slight departures from the background profile. Rays 7 and 8 that propagate in the Friedmann background do not feel the presence of the ERH.

\begin{figure}[h]
\begin{center}
\hspace{-7mm}
\includegraphics[scale=0.55]{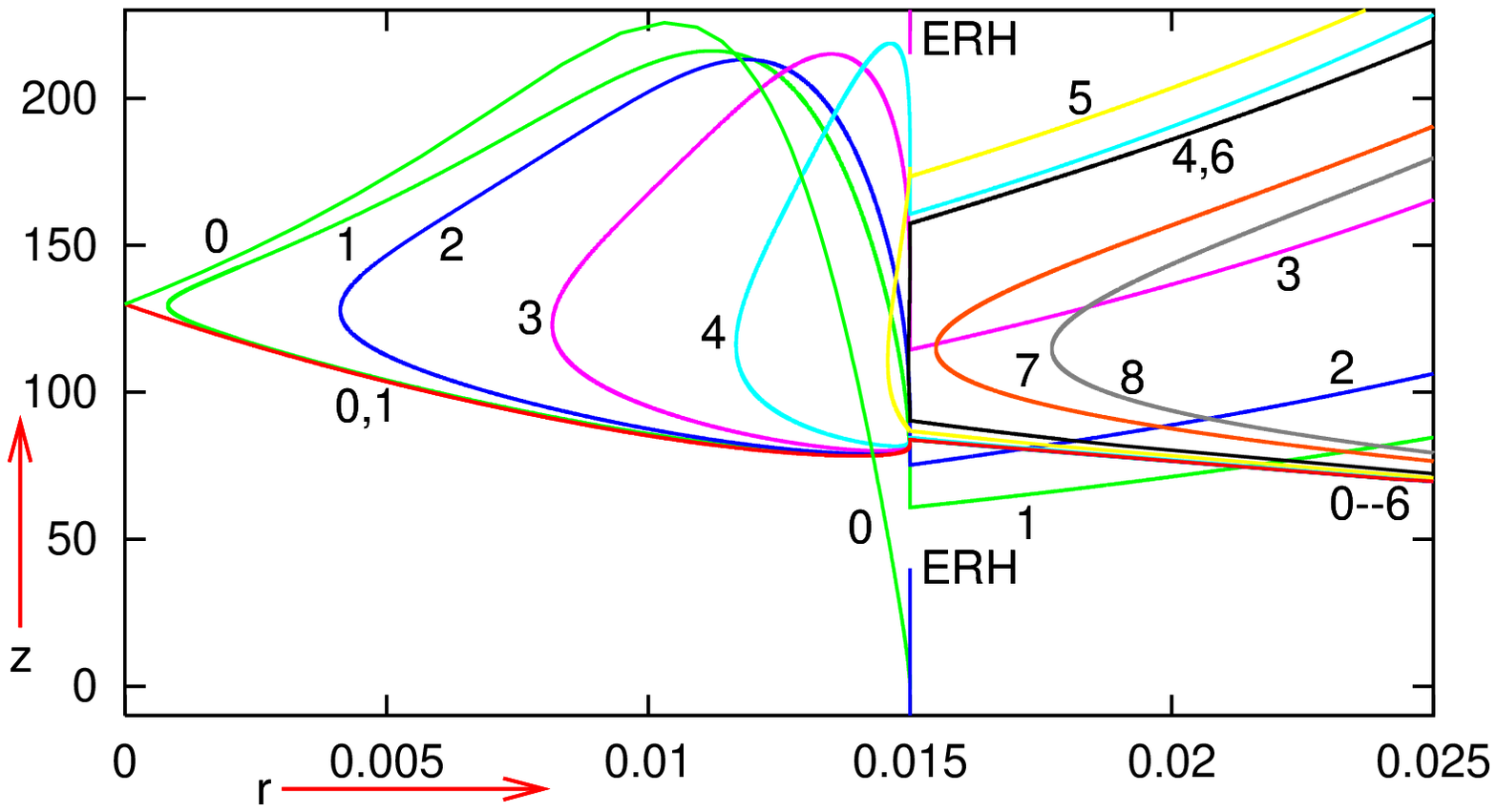}
${ }$ \\[-24mm]
\hspace{-55mm}
\includegraphics[scale=0.35]{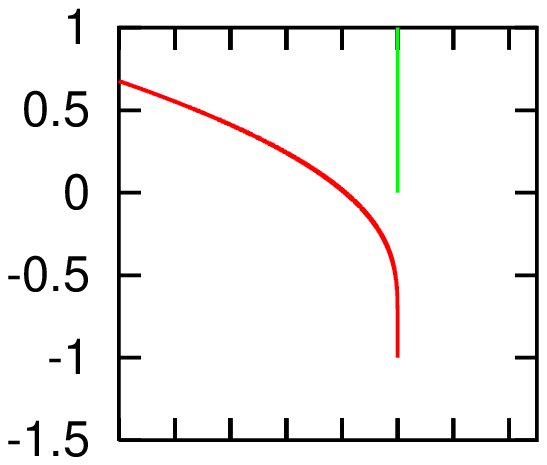}
${ }$ \\[10mm]
\hspace{-10mm}
\includegraphics[scale=0.55]{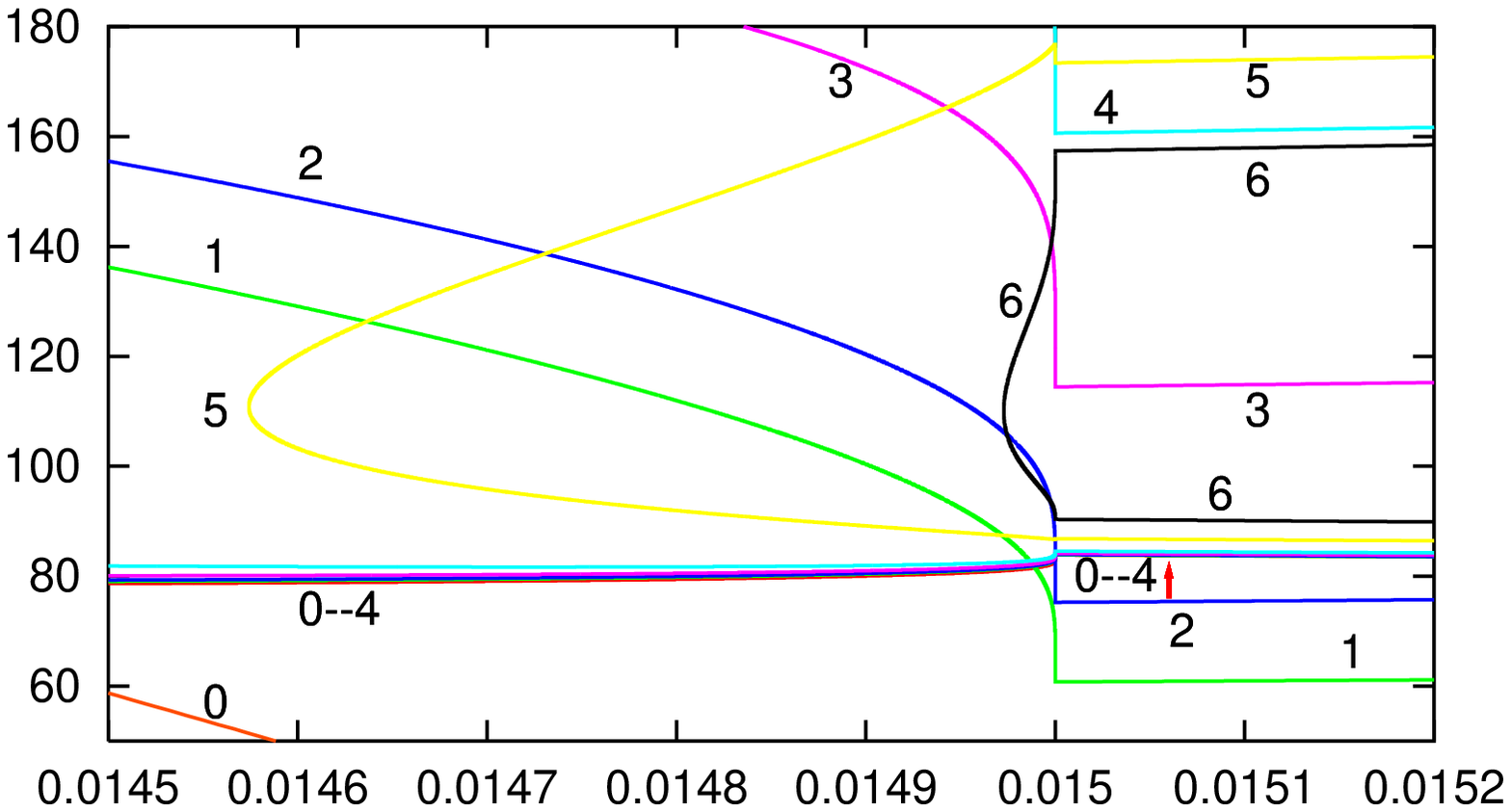}
\caption{{\bf Upper panel:} The $(z, r)$ relation along the rays from Table \ref{nonradrays}. The observer is at $r = r_{\rm O3}$ beyond the right
margin. Note the radial ray labelled ``0'' -- the only one that displays
blueshift to the observer. The inset shows that $z = -1$ at the BB. See more explanation in the text. {\bf Lower panel:} A closeup view of the knot in the upper panel.}
\label{drawzr}
\end{center}
\end{figure}

Figure \ref{drawzr} shows the $(z, r)$ relation in a neighbourhood of the value of $r \df r_{\rm ERH}$, at which the rays intersect the outer branch of the ERH. The observer is at $r = r_{\rm O3}$ beyond the right margin. This figure has to be read in the same way as Fig. \ref{drawzcent}, i.e., beginning at the observer and going in the direction of increasing affine parameter. Outside the ERH, $z$ increases with decreasing $r$ for $r_{\rm O3} > r > r_{\rm ERH}$, and increases with increasing $r$ when $r$ again becomes greater than $r_{\rm ERH}$, going to infinity at the intersection of the ray with the BB. The radial ray (labelled ``0'') is also included; it is the only one on which the observer sees $z < 0$ for a source at the LSH. The small inset in the main panel shows $z(r)$ on the radial ray near the BB, where $z \to -1$; the vertical stroke marks $r = r_{\rm ERH}$. The lower panel shows a further-magnified closeup on the near vicinity of $r = r_{\rm ERH}$.

These graphs demonstrate that the ERH, which was determined using only radial rays (see Sec. \ref{maxred} and Ref. \cite{Kras2014d}), is the locus of extrema of $z$ also on nonradial rays.

\begin{figure}[h]
\begin{center}
\hspace{-7mm}
\includegraphics[scale=0.55]{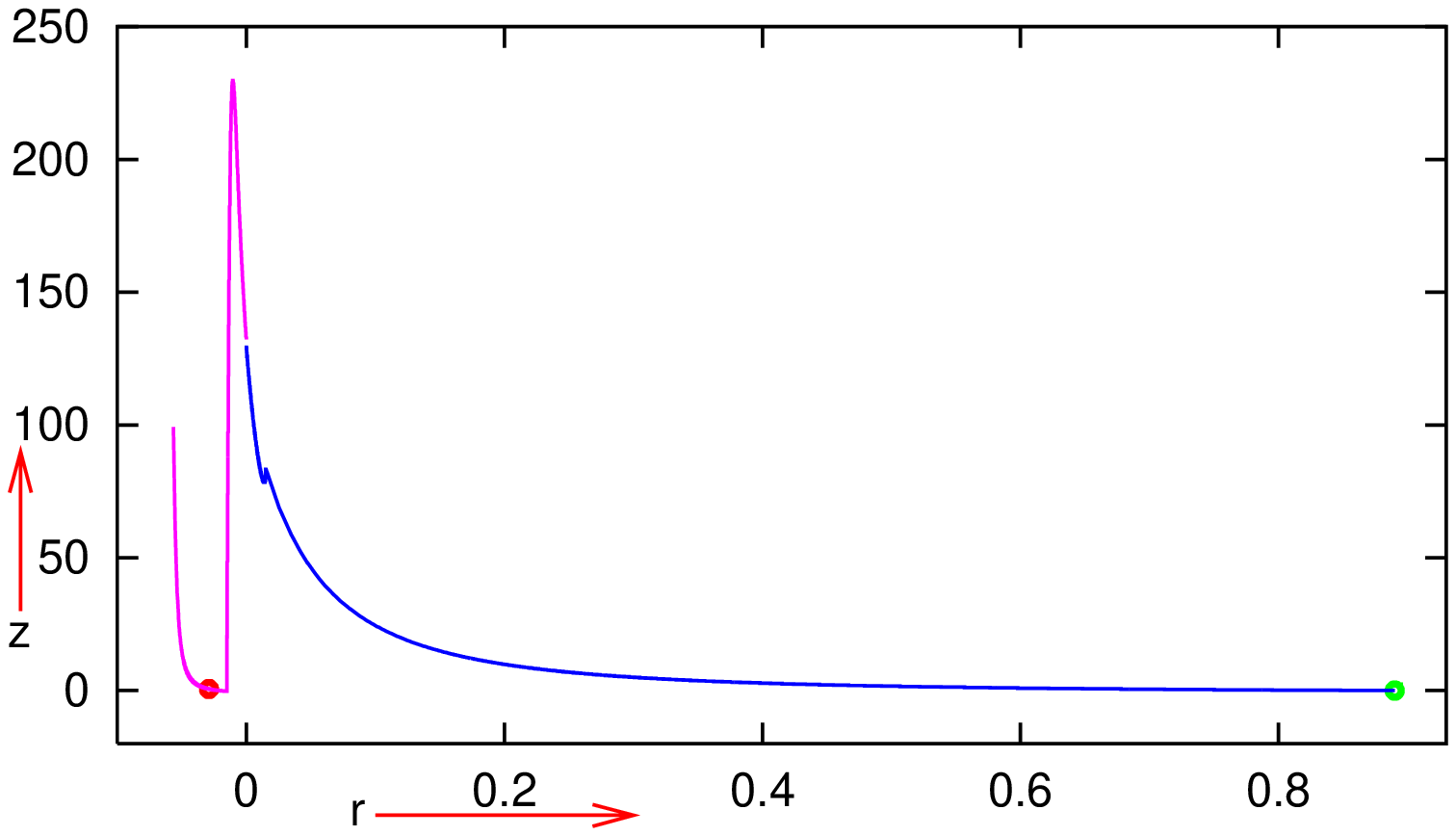}
${ }$ \\[-45mm]
\hspace{20mm}
\includegraphics[scale=0.55]{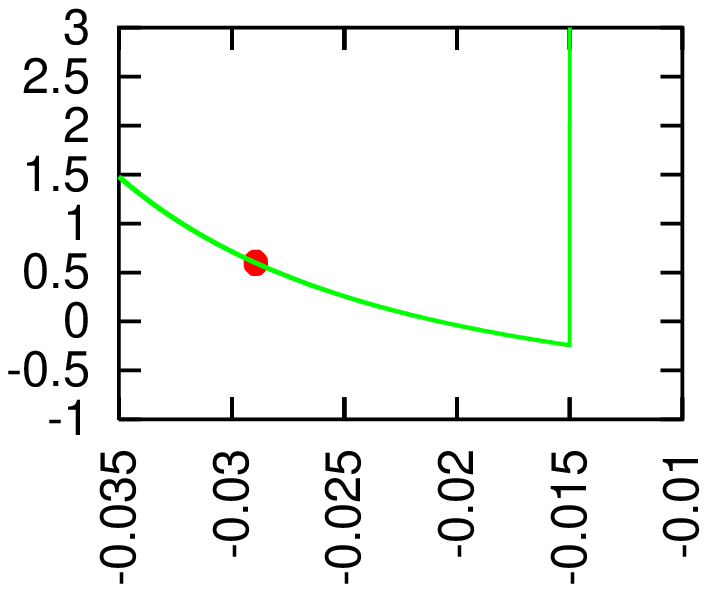}
\vspace{1cm} \caption{An illustration to the deception that may happen when
redshift is interpreted against a Friedmann background model. See text for
explanation. {\bf Main panel:} The full redshift profile between the BB (left
end of the curve) and the observer (right dot). The left dot marks the redshift
at the LSH. {\bf Inset:} A magnified view of the neighbourhood of the left dot
in the main panel.} \label{deceive}
\end{center}
\end{figure}

Finally, Fig. \ref{deceive} shows an example of a deception that may befall an
observer when she uses a Friedmann model to interpret the redshift that was
generated in an inhomogeneous Universe. This is the $z(r)$ graph along the
later of the two (radial) rays shown in Fig. \ref{drawafter}. The $r$-coordinate of observer O$_3$ is marked with the right dot in the main panel. The $r$-coordinate of the LSH is marked with the left dot. The redshift measured by O$_3$ for light emitted at the LSH is $z_d \approx 0.598$. When interpreted against the standard $\Lambda$CDM model with the parameters given in (\ref{7.8}), this leads to the conclusion that the ray was emitted $5.9 \times 10^9$ years ago \cite{wripa}. The observer would not suspect that between the LSH and herself (i.e., closer than the dot in the graph) there are sources of light for which the redshift would be much larger, up to $\sim 230$. In our model this ray was emitted at the LSH $0.13945$ NTU $= 1.3666 \times 10^{10}$ years ago. The redshift for a ray emitted that early in the $\Lambda$CDM model would be between 25 and 26 according to the same ``cosmology calculator'' \cite{wripa}.

\section{Accounting for the multitude of the GRBs}\label{multitude}

\setcounter{equation}{0}

So far, we discussed models of a single GRB. A model that would account for the multitude of observed GRBs should be imagined as a Friedmann background containing many humps like the ones in Figs. \ref{drawhighfreqray} or \ref{drawlowfreqray} of different shapes, spatial extents and heights above the flat part of $t_B(r)$, placed at different comoving positions.

For estimating how many GRB sources in the sky our model could accommodate, let us assume that all sources are described by Model 3. As explained in Sec. \ref{nonradial}, the GRB generator in Model 3 would fill the angular diameter of 2 degrees for observer O$_3$. Consider a unit sphere and a cone with the vertex at the center of the sphere that has the opening angle $2 \vartheta_0$. This cone subtends the solid angle $2 \pi (1 - \cos \vartheta_0)$. The full solid angle is $4 \pi$. So, counting naively, with $\vartheta_0 = 1^{\circ}$ = 0.01745 radians, the number of the corresponding solid angles in the full solid angle would be \begin{equation}\label{15.1}
\widetilde{\cal N}_{\gamma} = 2 / (1 - \cos \vartheta_0) = 13\ 131.
\end{equation}
However, to get a more realistic estimate we have to consider each circle on the unit sphere being inscribed into a square, and then calculate the ratio of the full solid angle to the area of that square. On a plane, the surface area ratio between a square of edge $a$ and a circle of diameter $a$ is $4/\pi$. So, the number above must be multiplied by $\pi/4$, and then the (approximately estimated) number of GRBs that could be visible in the sky {\it at the same time} comes out to be
\begin{equation}\label{15.2}
{\cal N}_{\gamma} = 10\ 313.
\end{equation}
This is probably too few to be realistic. However, with the angular size of the
GRB generator reduced by the factor $f$, the number ${\cal N}_{\gamma}$ would be
multiplied by $f^2$, and Model 3 needs such an improvement anyway.

\setcounter{equation}{0}

\section{Possible ways of improving the model}\label{improve}

\setcounter{equation}{0}

The influences on $1 + z$ of the various parameters of Models 1, 2 and 3 are interconnected. For example, decreasing $x_0$ alone (see Fig. \ref{drawpicture7}) has the immediate result of {\em in}creasing $1 + z$ (because the past-directed ray then hits the BB behind the hump and displays larger $z$ to the observer). A decrease in $1 + z$ is achieved when this change in $x_0$ is accompanied by decreasing $A_1$ (to make the steep straight segment longer) and decreasing $\Delta t_c$ (to make the ray hit the hump where it is steep, and at $t$ being as early as possible). But if the straight segment is too steep, then the ray and the LSH become nearly parallel and the numerical program has difficulty locating their intersection. In consequence of such inter-dependences between parameters, decreasing the height and width of the BB hump while keeping $1 + z$ sufficiently small is an extremely tedious process that requires a great number of long-lasting numerical experiments. (This is why the background model was left in an imperfect form, in which the value of $z$ at the LSH, eq. (\ref{8.5}), does not agree with (\ref{7.11}). This point needs to be improved, too.) Thus, the whole optimization should best be done by a computer program. Optimization by hand, applied in this paper, already allowed for a radical decrease in the size of the BB hump,\footnote{The crucial parameters are $A_0$, $B_0$ and $B_1$ that determine the height and width of the hump; $A_1$ is also important, but it is already small, see (\ref{9.19}). Obtained by the hand-optimization process, their values given in (\ref{9.19}) are much smaller than the first ones that produced $1 + z$ consistent with (\ref{9.5}); they were $(A_0, B_0, B_1) = (0.016, 0.01, 0.09)$.} but it is not known how much smaller the hump can still be made.

With a lower BB hump, also the duration of the GRB would be reduced, and the hump would be further away from the observer, automatically making the angular size of the GRB generator smaller. With narrower humps, more of them could be fitted into the full solid angle.

One way of improving the model is to put more parameters into the BB profile. However, some of the improvements were achieved by nontrivial means. For example, surprisingly good results were achieved by taking higher-degree curves instead of ellipses in the profile shown in Fig. \ref{drawpicture7}, see Appendix \ref{humpshape} for more on this. Computerized optimization could help also here.

Higher numerical accuracy will be needed for controlling the duration of the GRBs and of the afterglows in the model. The discontinuities reported in Sec. \ref{shortlive} are clearly related to insufficient accuracy. However, a higher accuracy will result in longer run-times of the programs, and this is one more reason why optimization by computer would be useful.

The observed durations of the GRBs are determined by the intensities of the gamma-ray flashes. The ways of calculating the intensities must still be found.

It would be desirable to have a hump profile whose ERH would not include a wall as high and wide as that in Models 1 -- 3; see Figs. \ref{drawhighfreqray} -- \ref{drawtruerays}, \ref{trueviewmagni} -- \ref{drawafter} and \ref{drawcross}. As can be seen from (\ref{6.2}) and (\ref{6.3}), $t$ on the ERH is large where $|\dril {t_B} r|$ is large. Thus, the ERH wall is narrower when the segment of large $|\dril {t_B} r|$ is shorter, and is lower when the value of $|\dril {t_B} r|$ is smaller. A lower and thinner wall would result in reducing the duration of the GRB, and could help in controlling the duration of the afterglow, both of which are too long-lasting in the current models. However, a smaller width of the ERH wall reduces the build-up time of the blueshift, and thus would increase the final $1 + z$. So, the difficulty here is in achieving a compromise between two contradictory goals: making the minimum of $1 + z$ on rays within the GRB sufficiently small, and the increase in $1 + z$ on rays within the afterglow sufficiently fast.

As already stated in Sec. \ref{collimation}, going over to the more general Szekeres metrics \cite{Szek1975,Szek1975b}, \cite{PlKr2006} may help in achieving the collimation of the GRB flashes. This is not a particularly pressing problem (the collimation is only a hypothesis \cite{gammainfo}), but employing the Szekeres model could also modify the results of Sec. \ref{nonradial} in interesting ways. For example, an anisotropic BB hump might result in reduced angular size of the GRB generator when viewed from certain directions.

Throughout this paper, an ambitious approach was taken, trying to explain the observed GRBs as if they all arose by blueshifting the relic radiation. This was to avoid the accusation that the author made his task unduly easy. Therefore, it was assumed that the lowest-frequency GRBs originate as the lowest-frequency emission radiation of hydrogen atoms, and correspondingly for the highest frequencies. But it is possible that there are several mechanisms of producing the GRBs, and that blueshifting from the last scattering is only one of them. In that case, one can also consider shifting high emission frequencies to low GRB frequencies. For example, the blueshifts needed to shift the $\nu_{\rm Hint}$ and the $\nu_{\rm Hmax}$ given by (\ref{9.3}) and (\ref{9.2}) to the minimum gamma-ray frequency given by (\ref{9.4}) are, respectively, $1 + z_{\rm int} \approx 9.1 \times 10^{-4}$ and $1 + z_{\rm opt} \approx 1.33 \times 10^{-3}$. These values would be much easier to achieve than those considered in this paper. It would suffice to make the hump lower by reducing $A_0$ and $B_0$, and then suitably reducing $\Delta t_c$. The result would be as described in paragraph 2 of this section.

\section{Conclusions}\label{conclu}

\setcounter{equation}{0}

The cosmological model employed here is a spatially homogeneous Friedmann background (see Sec. \ref{background}) into which a suitable number of Lema\^{\i}tre -- Tolman regions (see Secs. \ref{LTintro} and \ref{LTnullgeo}) is matched. Use was made of the long-known property of the L--T model that light rays emitted {\em radially} near those points of the Big Bang where the bang-time function is nonconstant ($\dril {t_B} r \neq 0$) display arbitrarily large blueshifts instead of redshifts \cite{Szek1980,HeLa1984,Kras2014d}. The present paper tested the hypothesis that the gamma-ray bursts (or at least some of them) arise by this mechanism. The blueshifts are generated only in the initial segments of the rays, before they exit the extremum-redshift hypersurface (see Sec. \ref{maxred}). In the further part of the rays' journey, redshifts are generated that can wipe out with excess the earlier-acquired blueshifts. The technical problem to solve was this: Can the parameters of the model be chosen so that a substantial part of the blueshift survives the journey to the present observer and the observed frequency of the ray is in the gamma-ray range? Section \ref{modelfit} answers this question in the positive.

The models discussed here satisfactorily reproduce the GRB frequency range (property (1) in Sec. \ref{GRBdata}, see Sec. \ref{modelfit}) and account for the large distances to the GRB sources (property (5), see Sec. \ref{distances}). However, if the mechanism of generating the GRBs is such as discussed here, then the distances reported in the literature are calculated by an incorrect method and are heavily underestimated, see Sec. \ref{distances}.

The other properties listed in Sec. \ref{GRBdata} are accounted for only qualitatively, with varying degrees of success. The duration of the GRBs (Property (2)) comes out too long, but Model 3 of Sec. \ref{modelfit} contains a free parameter that can control this quantity, see Sec. \ref{adaptdur}. To actually carry out the control, a much higher numerical accuracy would be necessary than was available to this author.

The afterglows (Property (3)) necessarily exist in these models, but they last longer than observed and no way of controlling their duration was provided (see Sec. \ref{afterglow}). As stated in the previous section, the observed durations are determined by the varying intensity of the radiation, which was not calculated in this paper.

In Sec. \ref{nonradial}, rays that are nonradial with respect to the L--T region were discussed. It was shown that the angular size of the GRB generator implied by Model 3 of Sec. \ref{modelfit} is about twice as large as the resolution of the GRB detectors. So, with moderate improvements in the model that are anyway needed for other purposes (see Sec. \ref{improve}), the perturbations in the CMB radiation implied by our model will not be larger than those actually caused by the GRBs.

The collimation of the GRBs into narrow jets (property (4)) appears to the observer as an illusion, as explained in Sec. \ref{collimation}. This property is not directly implied by observations, it is just a hypothesis. However, it may be possible to deal with it when the quasi-spherical Szekeres (QSS) metric \cite{Szek1975,Szek1975b}, \cite{PlKr2006} is employed instead of L--T.

In Sec. \ref{improve} the improvements needed in the model were discussed.

Even if the models presented here prove to be unsatisfactory explanations of the GRBs, they say new interesting things about the physics (in particular, optics) in the L--T models.

\appendix

\section{The age of the Universe at last scattering vs. redshift}\label{ageLS}

\setcounter{equation}{0}

There is a spurious discrepancy between the data on the age of the Universe at
last scattering, $\tau_{\rm LS}$, and the redshift of CMB. Namely, once $k = 0$ and $\Omega_{\Lambda} = 0.68$ have been specified for the $\Lambda$CDM model, the metric is uniquely defined, and then, given redshift,  $\tau_{\rm LS}$ can be computed or vice versa. The value of the redshift of the CMB radiation is currently given as $z_{\rm LS} = 1090$ \cite{Plan2014,Plan2014b}, and then the calculation based on the $\Lambda$CDM metric (\ref{7.4}) leads to $\tau_{\rm LS} = 4.77 \times 10^5$ y given by (\ref{7.16}). On the other hand, the most-often cited value is $\tau_{\rm LS} = 3.8 \times 10^5$ y \cite{swinxxxx} (which, by the same method, would imply ${\overline z}_{\rm LS} = 1269.3$).

The solution of this seeming contradiction is this: the $\Lambda$CDM model of
(\ref{7.4}) does not apply before the last scattering because it implies zero
pressure via the Einstein equations. For that epoch, the pressure of radiation
and matter cannot be neglected, and a more general model must be used, which, given $z_{\rm LS} = 1090$ implies a $\tau_{\rm LS}$ value close to that given above \cite{calculator}. The redshift at last scattering is calculated by sophisticated methods of particle physics -- see Ref. \cite{wikirecom} for a readable account.

\section{Blueshifting the black-body spectrum}\label{hotGRB}

\setcounter{equation}{0}

The maximum observed intensity of the CMB radiation by today is $I_{\rm max} \approx 5 \times 10^{-15}$ IU, where IU $\df$ W/(cm$^2 \times$ sr $\times$ Hz), at the frequency $\nu_{\rm max} \approx 2 \times 10^{11}$ Hz \cite{CMBspectrum}. Since $T/\nu_{\rm max}$ is constant (Wien's law), it follows from Planck's formula $I(\nu) = \frac {2 {\rm h}} {c^2} \times \frac {\nu^3} {{\rm exp}[{\rm h} \nu / ({\rm k} T)] - 1}$ that $I_{\rm max} \propto{\nu_{\rm max}}^3$. The radiation frequency obeys $\nu_{\rm emitted} = \nu_{\rm received} (1 + z)$, by the definition of $z$. The redshift between the instant of emission of the CMB radiation and the present time is $1 + z_{\rm LS} \approx 1091$ by (\ref{7.11}), so the maximum intensity of the relic radiation at the time of emission must have been $I_{\rm max} \times (1 + z_{\rm LS})^3 \approx 6.49 \times 10^{-6}$ IU. If this were blueshifted by $1 + z = 10^5$ to the gamma-ray frequency range, the resulting black-body radiation would have the temperature $3 \times 10^8$ K and peak intensity equal to $6.49 \times 10^9$ IU. This is many times more than observed -- exemplary GRBs have intensities below $10^{-24}$ W/(cm$^2$ Hz) \cite{McBr2006}. Thus, the GRBs cannot arise by blueshifting the thermal radiation with the black-body spectrum preserved -- and indeed the spectra of the GRBs are not thermal \cite{GRBrealspectra}.

\section{Remarks on choosing the shape of the BB hump}\label{humpshape}

\setcounter{equation}{0}

As the first Ansatz for the shape of the BB hump, the Gauss-type family of
curves was chosen
\begin{equation}\label{c.1}
t_B(r) = t_{\rm Bf} + A {\rm e}^{- B r^2},
\end{equation}
with $A$ and $B$ being adjustable constants and $t_{\rm Bf}$ given by
(\ref{8.3}). However, the two free parameters did not provide sufficient
flexibility. The rays (integrated backward in time) either hit the LSH with insufficient blueshift, or escaped back out through the ERH and acquired large \textit{red}shift at the LSH. The strongest blueshifts were of the order of $1 + z \approx 10^{-3}$ instead of the desired $10^{-5}$ or less.

The next shape that was tested was similar to that in Fig. \ref{drawpicture7}, except that the upper left curve was also an ellipse. With this, the range of blueshifts given by (\ref{9.5}) -- (\ref{9.8}) was achieved, but the humps were rather high and wide. In one model, which accounted for (\ref{9.7}), the time difference between the maximum of $t_B$ and its flat part was $0.21 T$, with $T$ given by (\ref{7.10}), and the comoving radius of the hump was $r_h = 0.09$, with the observer sitting at $r \sim 0.39$. In the other model the hump was only 3 times lower, and just as wide.

The shapes of the humps that were finally used (see Sec. \ref{modelfit}) emerged as the next corrections with respect to the one with two ellipses. They resulted in a radical reduction of both the height and width of the hump.

\section{Calculating $1 + z$ at the LSH}\label{numtricks}

\setcounter{equation}{0}

In the neighbourhood where the rays from Figs. \ref{drawhighfreqray} and
\ref{drawlowfreqray} cross the LSH, it lies extremely close to the BB. In Model
1 the difference in $t$ between them along the ray defined by (\ref{9.13}) is
$\sim 5.54 \times 10^{-13}$ NTU $\approx 5.43 \times 10^{-2}$ y $\approx 20$
days. Therefore, determining the instant when the ray crosses the LSH, and the value of $1 + z$ at that moment, requires a suitably small numerical step $\Delta \lambda$ in the affine parameter $\lambda$. With too large a step the integration jumps over the LSH, and crosses the BB without noting that it had crossed the LSH in the same step. A very small step along the whole ray would make the calculation prohibitively long. Sometimes the same problem appeared in a vicinity of other locations, for example, in crossing the ERH or in determining the instant of observation of a ray. In such cases, the step in $\lambda$ was changed in flight to either increase the precision near to critical locations, or relax it to accelerate the calculation.

For example, in Model 1, $\Delta \lambda$ was
\begin{equation}\label{d.1}
\Delta \lambda_1 = 2.5 \times 10^{-6},
\end{equation}
all along the ray displayed in Fig. \ref{drawhighfreqray}, which crossed the LSH in step \# 31,184,648, with
\begin{equation}\label{d.2}
\left(\begin{array}{c}
1 + z \\
t \\
\end{array}\right)_{\rm LSH1} = \left(\begin{array}{l}
 9.11448916340873438 \times 10^{-10} \\
 -0.13941448281939275\ {\rm NTU} \\
 \end{array}\right),
\end{equation}
and it crossed the BB in step \# 31,184,684 at
\begin{equation}\label{d.3}
t_{\rm BB1} = -0.13941448281944963\ {\rm NTU}
\end{equation}
with $1 + z$ at the level of $10^{-11}$. The exact value was not captured by the program (the last $1 + z$ that was reported was negative, i.e. arose
\textit{after} crossing the BB).

For Model 2, the increments in $\lambda$ were
\begin{equation}\label{d.4}
\Delta \lambda_2 = 2.5 \times 10^{-8}.
\end{equation}
The ray from Fig. \ref{drawlowfreqray} crossed the LSH in step \# 5,175,194 with
\begin{equation}\label{d.5}
\left(\begin{array}{c}
1 + z \\
t \\
\end{array}\right)_{\rm LSH2} = \left(\begin{array}{l}
 1.07858890707746014 \times 10^{-7} \\
 -0.13945300368712291\ {\rm NTU} \\
 \end{array}\right),
\end{equation}
and it crossed the BB in step \# 5,175,204 with
\begin{equation}\label{d.6}
t_{\rm BB2} = -0.13945300368714261\ {\rm NTU}
\end{equation}
and $1 + z \approx 5 \times 10^{-9}$. The problem with capturing the last $1 +
z$, mentioned under (\ref{d.3}), recurred also here.

\section{Problems with calculating $k^r$ from (\ref{4.13})}\label{negroot}

\setcounter{equation}{0}

Close to the BB, where $R \approx 0$, the expression under the square root in (\ref{4.13}) (call it $E_{413}$) becomes a difference of two very large numbers. This follows from (\ref{4.10}): where $R \to 0$ with $C \neq 0$, necessarily $(k^t)^2 \to \infty$. The $E_{413}$ must be positive, but sometimes the program could not ensure $E_{413} > 0$ at small $R$. In calculating the paths of the rays in Figs. \ref{illustr} and \ref{drawpolar} -- \ref{drawcross}, whenever $E_{413}$ was negative, the program would replace it by $|E_{413}|$, on the assumption that $E_{413} < 0$ resulted from numerical errors in calculating $E_{413} \approx 0$.

Along every ray in Figs. \ref{drawpolar} -- \ref{drawcross}, $k^r$ was expected
to change sign from $-$ to + at a certain point. In those cases, $E_{413} < 0$ was a signal that $k^r$ was near zero, and from that point on the program changed the sign of $k^r$.

\section{Deriving (\ref{12.6})}\label{der133}

\setcounter{equation}{0}

The equation of a radial null geodesic for (\ref{7.1}), using (\ref{7.2}) and
$r = 0$ at the observer O, has the solution
\begin{equation}\label{f.1}
\eta = \eta_{\rm O} - \ln \left(\sqrt{- k} r + \sqrt{1 - k r^2}\right).
\end{equation}
The value of $\eta_{\rm O}$ is calculated from (\ref{7.2}), taking (\ref{8.3})
for $t_B$ and $t = 0$ for the observer; it is
\begin{equation}\label{f.2}
\eta_{\rm O} = 0.59249644326164175.
\end{equation}
With $r = 0$ assumed for the observer, $r$ at D is the $r_{\rm O3}$ from
(\ref{10.7}), and then, from (\ref{f.1})
\begin{equation}\label{f.3}
\eta_D = 5.57539803067893525 \times 10^{-2}.
\end{equation}
Using this in (\ref{7.2}) we find
\begin{eqnarray}
t_D &=& t_B + \frac {M_0} {(-k)^{3/2}} \left(\sinh \eta_D - \eta_D\right)
\nonumber \\
&=& 1.14196789587062714 \times 10^{-4}\ {\rm NTU}, \label{f.4} \\
S_D &=& 3.886639547232007 \times 10^{-3},
\end{eqnarray}
and from here (\ref{12.6}) follows.

{\bf Acknowledgements.} I am grateful to Krzysztof Bolej\-ko for a helpful
discussion on last scattering and to Linda Sparke from NASA for several bits of
information about the properties of the GRBs and their detection techniques.

\bigskip

\end{document}